\documentclass[manuscript]{aastex}

\usepackage{amsmath}

\newcommand{\pdn}[3]{\dfrac{\partial^{#3} #1}{\partial #2^{#3}}}
\newcommand{\rd}{\mathrm{d}}

\def\pmb#1{\mbox{\boldmath$#1$}}
\def\pmbmt#1{\pmb{\sf #1}}

\def\gtsim {>\kern-1.2em\lower1.1ex\hbox{$\sim$}}
\def\ltsim {<\kern-1.2em\lower1.1ex\hbox{$\sim$}}

\slugcomment{Not to appear in Nonlearned J., 45.}

\shorttitle{Magnetic discrete modes}
\shortauthors{H.Asai and U.Lee}

\begin{document}


\title{Axisymmetric toroidal modes of general relativistic magnetized
  neutron star models}


\author{Hidetaka \textsc{Asai}}
\affil{Tohoku University, Sendai, Miyagi
980-8578}
\email{hasai@astr.tohoku.ac.jp}
\and
\author{Umin \textsc{Lee}}
\affil{Astronomical Institute, Tohoku University, Sendai, Miyagi 980-8578}
\email{lee@astr.tohoku.ac.jp}



\begin{abstract}
We calculate axisymmetric toroidal modes of magnetized neutron stars
with a solid crust in the general relativistic Cowling approximation. 
We assume that the interior of the star is threaded by a poloidal magnetic field, which is continuous at the surface with an outside dipole field.
We examine the cases of the field strength $B_{{\rm{S}}}\sim10^{16}$ G at the surface. 
Since separation of variables is not possible for the oscillations of
magnetized stars, we employ finite series expansions for the
perturbations using spherical harmonic functions.
We find discrete normal toroidal modes of odd parity, 
but no toroidal modes of even parity are found.
The frequencies of the toroidal modes 
form distinct mode sequences and the frequency in a given mode sequence
gradually decreases as the number of radial nodes of the eigenfunction
increases. 
From the frequency spectra computed for neutron stars of different masses, we find that the frequency is
almost exactly proportional to $B_{{\rm{S}}}$ and is well
represented by a linear function of $R/M$ for a given 
$B_{{\rm{S}}}$,  
where $M$ and $R$ are the mass and radius of the
star. 
The toroidal mode frequencies for $B_{{\rm{S}}}\sim  10^{15}$
G are in the
frequency range of the quasi-periodic oscillations (QPOs) detected in the soft-gamma-ray repeaters, but 
we find that 
the toroidal normal modes cannot explain all the detected QPO frequencies.
\end{abstract}





\section{Introduction}

Recent discovery of quasi-periodic oscillations (QPOs) in the decaying tail of the hyper-flares of the three magnetar candidates  
(hyper-flares from SGR 0526-66 observed in 1979, from SGR 1900+14 in 1998, and from SGR 1806-20 in 2004) has triggered intensive
studies of the oscillations of magnetized neutron stars.
Observed QPO frequencies are $\sim$18, $\sim$30, and 92.5Hz for SGR 1806-20 hyperflare (Israel et al. 2005), $\sim$28, 53.5, 84, and 155Hz for SGR 1900+14 (Strohmayer \& Watts 2005). 
Additional frequencies of 150 and 626Hz have been identified for SGR
1806-20 (Strohmayer \& Watts 2006), and a recent reanalysis for SGR
1806-20 by Hambaryan et al. (2011) indicates
QPOs at 16.9, 21.4, 36.8, 59.0, 61.3, and 116.3 Hz.
The QPOs are now interpreted as a manifestation of global oscillations of the underlying neutron stars, and 
it is expected that we will be able to carry out seismological studies of magnetars, neutron stars with an extremely strong magnetic field.

Since the discovery of the QPOs in the magnetar candidates, many researchers have suggested that the QPOs are caused by the oscillations of the underlying neutron star, especially by the crustal torsional modes of the star.
For example, it has been suggested that the QPOs are generated by seismic vibrations of the neutron star crust (Israel et al. 2005) and that the QPOs are mainly due to low $l$ fundamental toroidal modes excited in the crust (Strohmayer \& Watts 2005, 2006). 
One of the reasons for such suggestions is that the torsional mode
frequencies overlap the observed QPOs and the toroidal crust modes confined in the solid crust might be easily excited by a giant flare, which is probably associated with a restructuring of the
magnetic field (Strohmayer \& Watts 2006).

Since magnetars are believed to possess an extremely strong magnetic field in the interior, 
to determine the oscillation frequency spectra of the stars,
we need to correctly take account of the effects of the strong magnetic field on the oscillations, that is,
we have to calculate not only the crustal toroidal modes modified by the magnetic field, but also the oscillation modes in the fluid core threaded by the magnetic field.
Theoretical calculations of axisymmetric toroidal modes of strongly magnetized neutron stars have so far been carried out by many authors
in Newtonian formulation (e.g., Piro 2005; Glampedakis, Samuelsson \&
Andersson 2006; Levin 2007; Lee 2007, 2008; van Hoven \& Levin 2011) and
in general relativistic formulation (e.g., Messios et al. 2001; 
Sotani, Kokkotas \& Stergioulas 2007, 2008; Colaiuda \& Kokkotas 2011,
2012;
Gabler et al. 2011, 2012, 2013; van Hoven \& Levin 2012). 
Since the shear
modulus may be dominated by the magnetic pressure in most parts of the crust for 
a dipole field of strength $B\gtsim 10^{15}$ G, 
the crustal toroidal modes are not necessarily well decoupled from the oscillation modes in the fluid core
when the core is also threaded by the magnetic field. 
If the toroidal modes in the crust are strongly coupled with magnetic modes
in the core, we have to consider oscillation modes propagating in the entire region of the star.
Several authors have calculated such axisymmetric magnetic oscillations using
magnetohydrodynamical (MHD) simulations (e.g., Sotani, Kokkotas \&
Stergioulas 2008, Colaiuda \& Kokkotas
2011, 2012; Gabler et al. 2011, 2012, 2013). 
Most of the authors using MHD simulations to study the oscillations of magnetized stars
are able to show the existence of QPOs associated with
the edges of the Alfv\'en continua in the core and are doubtful about the existence of
discrete normal toroidal modes in strongly magnetized neutron stars.
However, for example, Colaiuda \& Kokkotas (2011) have suggested that  
a crustal toroidal mode can survive as a discrete mode 
if the frequency is in a gap between the Alfv\'en continua in the core
(see aslo Levin 2007; van Hoven \& Levin 2011). 
We think it still worthwhile to investigate the oscillations of strongly magnetized neutron stars
using a method of calculation different from MHD simulations.

In this paper, we look for axisymmetric normal toroidal modes of magnetized neutron stars with a solid crust
using the general relativistic formulation given in Sotani, Kokkotas \& Stergioulas (2007).
The numerical method to compute normal modes of magnetized neutron stars is the same as that in Lee (2008), 
who employed series expansions of a finite length in terms of spherical harmonic functions 
for the perturbations.
We solve two different sets of the oscillation equations, one for fluid regions and the other for a solid crust, and match the solutions at the interfaces between the crust and fluid regions to obtain a complete solution for a mode.
We calculate frequency spectra of the axisymmetric toroidal modes of neutron stars threaded by a dipole magnetic field, and compare the computed frequencies with the observed QPO
frequencies. 
\S 2 describes the method used to construct a neutron star model and a poloidal magnetic field,
and the perturbation equations for axisymmetric toroidal modes in magnetized stars are derived in \S 3.
Numerical results are summarized in \S 4. 
Discussions are given in \S 5 and 
we conclude in \S 6.
The details of the oscillation equations solved in this paper are given in the Appendix.
Unless otherwise noted, we adopt units of $c=G=1$, where $c$ and $G$ denote the speed of light and the gravitational constant, respectively, and the metric signature is $\left(-,+,+,+\right)$.

\begin{table*}
\begin{center}
\caption{The main properties of seven neutron star models}
\label{symbols}
\begin{tabular}{@{}lccccc}
\hline
Model & $M/M_\odot$ & $R~({\rm km})$
        & $\rho_c\left({\rm{g \ cm^{-3}}}\right)$
        & crust thickness $\left({\rm{km}}\right)$ & $R/M$ \\
\hline
NS05 & 0.499 & 12.235 & $4.771\times 10^{14}$ & 3.102 & 16.584 \\
NS08 & 0.799 & 11.940 & $6.260\times
  10^{14}$ & 1.887 & 10.111 \\
NS10 & 0.999 & 11.882 & $7.303\times 10^{14}$ & 1.449 & 8.042 \\
NS12 & 1.198 & 11.823 & $8.446\times 10^{14}$ & 1.151 & 6.680 \\
NS14 & 1.394 & 11.729 & $9.788\times
  10^{14}$ & 0.916 & 5.693 \\
NS16 & 1.599 & 11.564 & $1.159\times 10^{15}$ & 0.722 & 4.896 \\
NS18 & 1.797 & 11.284 & $1.419\times 10^{15}$ & 0.551 & 4.249 \\
\hline
\end{tabular}
\medskip
\end{center}
\end{table*}

\section{Equilibrium configuration}

We assume that the neutron star, which is non-rotating and threaded by a dipole magnetic field, is static and spherically symmetric,
ignoring the possible deformation of the star due to the magnetic field because the gravitational energy is much larger than the magnetic energy. 
To calculate neutron star models, we integrate the Tolman-Oppenheimer-Volkoff (TOV) equation using appropriate
equations of state, and if we write the line element as
\begin{eqnarray}
{\rm{d}} s^2=g_{\alpha\beta}{\rm{d}} x^\alpha{\rm{d}} x^\beta=-e^{2\Phi}{\rm{d}} t^2+e^{2\Lambda}{\rm{d}} r^2+r^2\left({\rm{d}}\theta^2+\sin^2\theta{\rm{d}}\phi^2\right),
\end{eqnarray}
the functions $\Phi$ and $\Lambda$ satisfy the following equations
\begin{eqnarray}
e^{-2\Lambda}=1-\frac{2M_r}{r}, \quad
\frac{{\rm{d}}\Phi}{{\rm{d}}r}=\frac{M_r+4\pi r^3p}{r^2\left(1-2M_r/r\right)}, 
\end{eqnarray}
where $M_r=\int_0^r4\pi r^2\rho{\rm{d}} r$ is the gravitational mass, $p$ is the pressure, and $\rho$ is the energy density of the star.

We use the equation of state, for the inner crust and the fluid core, given by Douchin \& Haensel (2001), who assume $npe\mu$ matter in the neutron star core, and 
the equation of state given by Baym, Pethick \& Sutherland (1971) for the outer crust.
For the fluid ocean we simply use the equation of state for a mixture of a completely degenerate electron gas and a non-degenerate gas of Fe nuclei.
For the solid crust, we employ the average shear modulus $\mu_{{\rm{eff}}}$ (Strohmayer et al. 1991), which in the limit of $\Gamma\equiv\left(Ze\right)^2/\left(ak_{\rm{B}}T\right)\rightarrow\infty$ is given by
\begin{eqnarray}
\mu_{{\rm{eff}}}=0.1194\frac{\left(Ze\right)^2n}{a},
\end{eqnarray}
where $n$ is the number density of the nuclei and $a$ is the separation between the nuclei defined by
\begin{eqnarray}
\frac{4\pi}{3}a^3n=1,
\end{eqnarray}
and $k_{\rm{B}}$ is the Boltzmann constant.
We use seven neutron star models, which have different masses and central
densities.
In Table 1, we summarize the physical properties of the seven neutron star models.

For the magnetic field, we assume an axisymmetric poloidal field produced by a toroidal 4-current $J_{\mu}$.
If we give the 4-current $J_{\mu}$ and the electromagnetic 4-potential $A_{\mu}$ as
\begin{eqnarray}
J_{\mu}=\left(0,0,0,J_{\phi}\right), \quad A_{\mu}=\left(0,0,0,A_{\phi}\right),
\end{eqnarray}
the Maxwell's equation $F^{\mu\nu}_{\ ;\nu}=4\pi J^{\mu}$, where $F_{\mu\nu}=\partial_{\mu}A_{\nu}-\partial_{\nu}A_{\mu}$, is reduced to
\begin{eqnarray}
e^{-2\Lambda}\frac{\partial^2A_{\phi}}{\partial
  r^2}+\frac{1}{r^2}\frac{\partial^2
  A_{\phi}}{\partial\theta^2}+\left(\frac{{\rm{d}}\Phi}{{\rm{d}}r}-\frac{{\rm{d}}\Lambda}{{\rm{d}}r}\right)e^{-2\Lambda}\frac{\partial
  A_{\phi}}{\partial
  r}-\frac{1}{r^2}\frac{\cos\theta}{\sin\theta}\frac{\partial A_{\phi}}{\partial\theta}=-4\pi J_{\phi}.
\end{eqnarray}
Expanding the potential $A_{\phi}$ and the current $J_{\phi}$ as
\begin{eqnarray}
A_{\phi}\left(r,\theta\right)=a_{\ell}\left(r\right)\sin\theta\frac{\partial}{\partial\theta}P_{\ell}\left(\cos\theta\right),
\end{eqnarray}
\begin{eqnarray}
J_{\phi}\left(r,\theta\right)=j_{\ell}\left(r\right)\sin\theta\frac{\partial}{\partial\theta}P_{\ell}\left(\cos\theta\right), 
\end{eqnarray}
where 
$P_{\ell}\left(\cos\theta\right)$ is the Legendre polynomial of order $\ell$, 
we obtain the Grad-Shafranov equation associated with the order $\ell$:
\begin{eqnarray}
e^{-2\Lambda}\frac{{\rm{d}}^2a_{\ell}}{{\rm{d}}r^2}+\left(\frac{{\rm{d}}\Phi}{{\rm{d}}r}-\frac{{\rm{d}}\Lambda}{{\rm{d}}r}\right)e^{-2\Lambda}\frac{{\rm{d}}a_{\ell}}{{\rm{d}}r}
-\frac{\ell\left(\ell+1\right)}{r^2}a_{\ell}=-4\pi j_{\ell}.
\end{eqnarray}
For a dipole field, we use $\ell=1$, and
for the current $j_1$, we assume
\begin{eqnarray}
j_1=f_0r^2\left(\rho+p\right),
\end{eqnarray}
where $f_0$ is an arbitrary constant (Konno et al. 1999).
At the centre we may expand the function $a_1$ as
\begin{eqnarray}
a_1\approx\alpha_0r^2+{\cal{O}}\left(r^4\right),
\end{eqnarray}
where $\alpha_0$ is another arbitrary constant. 

Assuming $j_1^{\left({\rm{ex}}\right)}=0$ for a dipole magnetic field in the exterior region, $a_1^{(\rm ex)}$ is given as
\begin{eqnarray}
a_1^{\left({\rm{ex}}\right)}=-\frac{3\mu_b}{8M^3}r^2\left[\ln\left(1-\frac{2M}{r}\right)+\frac{2M}{r}+\frac{2M^2}{r^2}\right],
\end{eqnarray}
where $\mu_b$ is the magnetic dipole moment and $M$ is the mass of the star.
We determine the constants $\alpha_0$ and $f_0$ so that the interior solutions $a_1$ and ${\rm{d}} a_1/{\rm{d}} r$ are matched with $a_1^{\rm (ex)}$ and 
${\rm{d}} a_1^{\rm (ex)}/{\rm{
d}} r$ at the surface of the star.
The magnetic field $H_{\mu}=B_{\mu}/\sqrt{4\pi}$ is then given by
\begin{eqnarray}
H_r=\frac{e^{\Lambda}\cos\theta}{\sqrt{\pi}r^2}a_1,
\end{eqnarray}
\begin{eqnarray}
H_{\theta}=-\frac{e^{-\Lambda}\sin\theta}{\sqrt{4\pi}}\frac{{\rm{d}}a_1}{{\rm{d}}r},
\end{eqnarray}
and $H_t=H_{\phi}=0$, where $B_\mu=\frac{1}{2}\epsilon_{\mu\nu\alpha\beta}u^\nu
F^{\alpha\beta}$. $\epsilon_{\mu\nu\alpha\beta}$ is the completely
antisymmetric tensor with $\epsilon_{tr\theta\phi}=\sqrt{-g}$, where
$g$ is the determinant of $g_{\mu\nu}$.

\section{Method of solution}

The stress-energy tensor $T^{\mu\nu}$ for a magnetized neutron star
with a solid crust is given by the sum of contributions from a perfect fluid $T^{\mu\nu}_{{\rm{fluid}}}$, a magnetic field $T^{\mu\nu}_{{\rm{mag}}}$ and a shear stress $T^{\mu\nu}_{{\rm{shear}}}$, that is
\begin{eqnarray}
T^{\mu\nu}=T^{\mu\nu}_{{\rm{fluid}}}+T^{\mu\nu}_{{\rm{mag}}}+T^{\mu\nu}_{{\rm{shear}}},
\end{eqnarray}
where
\begin{eqnarray}
T^{\mu\nu}_{{\rm{fluid}}}=\left(\epsilon+p\right)u^{\mu}u^{\nu}+pg^{\mu\nu},
\end{eqnarray}
\begin{eqnarray}
T^{\mu\nu}_{{\rm{mag}}}=H^2u^{\mu}u^{\nu}+\frac{1}{2}H^2g^{\mu\nu}-H^{\mu}H^{\nu},
\end{eqnarray}
and $\epsilon\equiv\rho$ is the energy density of the matter, $H^2=H_\mu H^\mu$, and $u^\mu$ is the 4-velocity.

The equations of motion for the perturbations are given by the perturbed momentum conservation equation:
\begin{eqnarray}
\delta\left(h^{\mu}_{\ \alpha}T^{\alpha\nu}_{\ ;\nu}\right)=0,
\end{eqnarray}
where $\delta$ means Eulerian perturbation, and $h^{\mu}_{\ \nu}=g^{\mu}_{\
  \nu}+u^{\mu}u_{\nu}$ is the projection tensor, which projects the conservation of the energy-momentum on to the hypersurface normal to $u^{\mu}$.
The 4-velocity $u^{\mu}$ of the matter in the static equilibrium is given by
$
u^{\mu}=\left(e^{-\Phi},0,0,0\right).
$

For axisymmetric toroidal modes, for which
we assume $\delta u^t=\delta u^r=\delta u^{\theta}=0$ and $\delta\epsilon=\delta p=0$, 
equation (18) becomes
\begin{eqnarray}
\left(\epsilon+p+H^2\right)e^{-\Phi}{\partial\over\partial t}\delta u^{\phi}=\delta
H^{\phi}\bigg[\left(\frac{{\rm{d}}\Phi}{{\rm{d}}r}+\frac{2}{r}\right)H^r
+2\frac{\cos\theta}{\sin\theta}
  H^{\theta}\bigg] \nonumber \\
+H^r
{\partial\over\partial r}\delta
H^{\phi}+H^{\theta}{\partial\over\partial\theta}\delta H^{\phi}
-{\partial\over\partial r}\delta T^{r\phi\left({\rm{shear}}\right)}
-{\partial\over\partial\theta}\delta
T^{\theta\phi\left({\rm{shear}}\right)} \nonumber \\
-\left(\frac{4}{r}+\frac{{\rm{d}}\Phi}{{\rm{d}}r}+\frac{{\rm{d}}\Lambda}{{\rm{d}}r}\right)\delta
T^{r\phi\left({\rm{shear}}\right)}
-3\frac{\cos\theta}{\sin\theta}\delta
T^{\theta\phi\left({\rm{shear}}\right)}, \ 
\end{eqnarray}
where $\delta u^{\phi}=e^{-\Phi}\partial\xi^{\phi}/\partial t$ and $\xi^\phi$ is the $\phi$ component of the displacement vector $
\xi^\mu$.
Here, we have adopted the relativistic Cowling approximation, that is, $\delta g_{\mu\nu}=0$. 
According to Schumaker \& Thorne (1983), the perturbed shear stress tensors for the toroidal modes are given by
\begin{eqnarray}
\delta T_{r\phi}^{({\rm{shear}})}=-\mu
r^2\sin^2\theta\frac{\partial}{\partial
  r}\xi^{\phi}\left(t,r,\theta\right),\\
\delta T_{\theta\phi}^{({\rm{shear}})}=-\mu r^2\sin^2\theta\frac{\partial}{\partial\theta}\xi^{\phi}\left(t,r,\theta\right),
\end{eqnarray}
where $\mu$ is the shear modulus.

Using the Maxwell's equation $F_{\left[\mu\nu  ;\gamma\right]}=0$ under the ideal MHD approximation, we obtain the relativistic induction equation given by (e.g., Sotani, Kokkotas \& Stergioulas 2007)
\begin{eqnarray}
H^{\mu}_{\ ;\nu}u^{\nu}=-u^{\alpha}_{\ ;\alpha}H^{\mu}+u^{\mu}_{\
  ;\nu}H^{\nu}+H^{\alpha}u_{\alpha ;\beta}u^{\beta}u^{\mu}.
\end{eqnarray}
Taking Eulerian perturbations of equation (22), we obtain the perturbed induction equation for axisymmetric toroidal modes:
\begin{eqnarray}
e^{-\Phi}{\partial\over\partial t}\delta H^{\phi}={{\rm{d}}\Phi\over {\rm{d}} r}H^r\delta u^{\phi}+H^r{\partial\over\partial r}\delta
u^{\phi}+H^{\theta}{\partial\over\partial\theta}\delta u^{\phi}.
\end{eqnarray}

Substituting the magnetic field components (13) and (14) and the perturbed shear stress tensors (20) and (21) into equations (19) and (23), 
we obtain the perturbation equations of axisymmetric toroidal modes of the magnetized neutron stars:
\begin{eqnarray}
-e^{-2\Phi}\omega^2\xi^{\phi}\left[\epsilon+p+\frac{a_1^2}{\pi
    r^4}\left(\cos^2\theta+\eta\sin^2\theta\right)\right]
=\frac{e^{-\Lambda}}{\sqrt{\pi}r^2}\Bigg\{\Bigg[\left(\frac{{\rm{d}}\Phi}{{\rm{d}}r}+\frac{2}{r}\right)a_1
\ \ \ \ \ \ \ \ \ \ \ \ \nonumber \\
-\frac{{\rm{d}}a_1}{{\rm{d}}r}\Bigg]\cos\theta\delta
H^{\phi}
+a_1\cos\theta{\partial \over\partial r}\delta
H^{\phi}
-\frac{1}{2}\frac{{\rm{d}}a_1}{{\rm{d}}r}\sin\theta{\partial\over\partial\theta}\delta
H^{\phi}\Bigg\}
+{\partial\over\partial r}\left(\mu
  e^{-2\Lambda}{\partial\over\partial r}\xi^{\phi}\right) \nonumber \\
+\mu
e^{-2\Lambda}\left(\frac{4}{r}+\frac{{\rm{d}}\Phi}{{\rm{d}}r}+\frac{{\rm{d}}\Lambda}{{\rm{d}}r}\right){\partial\over\partial
  r}\xi^{\phi}
+\frac{\mu}{r^2}{\partial^2\over\partial\theta^2}\xi^{\phi}+\frac{3\mu}{r^2}\frac{\cos\theta}{\sin\theta}
{\partial\over\partial\theta}\xi^{\phi}, \ \ 
\end{eqnarray}
\begin{eqnarray}
\delta H^{\phi}=\frac{e^{-\Lambda}\cos\theta}{\sqrt{\pi}r^2}a_1{\partial\over\partial r}\xi^{\phi}-\frac{e^{-\Lambda}\sin\theta}{\sqrt{4\pi}r^2}\frac{{\rm{d}}a_1}{{\rm{d}}r}{\partial\over\partial\theta}\xi^{\phi},
\end{eqnarray}
where $\eta$ is defined by
\begin{eqnarray}
\eta\equiv\frac{1}{4}e^{-2\Lambda}\left(\frac{{\rm{d}}\ln a_1}{{\rm{d}}\ln r}\right)^2,
\end{eqnarray}
and we have assumed $\xi^{\phi}\left(t,r,\theta\right)=\xi^{\phi}\left(r,\theta\right)e^{{\rm{i}}\omega
  t}$ and $\delta H^{\phi}\left(t,r,\theta\right)=\delta
H^{\phi}\left(r,\theta\right)e^{{\rm{i}}\omega t}$.

For the perturbations of magnetized stars the separation of variables
between $r$ and $(\theta,\phi)$ is impossible, so we employ finite
series expansions of the axisymmetric 
perturbations in terms of
the spherical harmonic functions $Y_l^m\left(\theta,\phi\right)$, that is, 
\begin{eqnarray}
\xi^{\phi}\left(r,\theta\right)=\sum_{j=1}^{j_{{\rm{max}}}}T_{l_j'}\left(r\right)\frac{1}{\sin\theta}\frac{\partial}{\partial\theta}Y_{l_j'}^{m=0}\left(\theta,\phi\right),\\
\delta H^{\phi}\left(r,\theta\right)=\sum_{j=1}^{j_{\rm max}}b_{l_j}\left(r\right)\frac{1}{\sin\theta}\frac{\partial}{\partial\theta}Y_{l_j}^{m=0}\left(\theta,\phi\right),
\end{eqnarray}
where $l_j=2j$ and $l_j'=l_j-1$ for even modes, and $l_j=2j-1$ and $l_j'=l_j+1$ for odd modes for $j=1,2,3,\cdots,j_{{\rm{max}}}$  (Lee 2008). 
Most of the numerical results shown below are obtained for
$j_{{\rm{max}}}=15$. In this convention, the angular pattern of $\delta H^{\phi}$ (of $\xi^{\phi}$) at the stellar surface is symmetric (antisymmetric) about the equator for even modes and it is antisymmetric (symmetric) for odd modes.
Note that when the displacement vector is given as the sum of the spheroidal component and toroidal component (e.g., Unno et al 1989), the two components are decoupled for axisymmetric modes of non-rotating stars magnetized with a dipole field as given by equations (27) and (28), and they are coupled for non-axisymmetric modes (e.g., Lee 2007), and that
for rotating stars the two components are always coupled for both axisymmetric and non-axisymmetric modes.
Substituting the expansions into the perturbed equations, we obtain a set of linear ordinary differential equations, which
is formally written as
\begin{eqnarray}
r{{\rm{d}}\over {\rm{d}} r}\pmb{Y}=\pmbmt{C}\pmb{Y},
\end{eqnarray}
where
\begin{eqnarray}
\pmb{Y}=
\left(
\begin{array}{c}
\pmb{t}\\ 
\pmb{h} 
\end{array}
\right),
\end{eqnarray}
and $\pmbmt{C}$ is the coefficient matrix, and $\pmb{t}=(T_{l^\prime_j})$ and $\pmb{h}=(h_{l_j})$ with $h_{l_j}=r^3b_{l_j}/\left(\sqrt{\pi}a_1\right)$.
Note that the coefficient matrix $\pmbmt{C}$ and the dependent vector $\pmb{Y}$ are defined differently in the fluid 
and solid regions. See the Appendix for the detail.

The set of ordinary differential equations may be solved as an eigenvalue problem of $\omega$ by applying boundary conditions at the center and surface
of the star and jump conditions at the core-crust and crust-ocean interfaces. 
For the inner boundary condition, we require that the functions $\mbox{\boldmath $t$}$ and $\mbox{\boldmath $h$}$ are regular at the center. 
The outer boundary conditions are given by $\mbox{\boldmath $h$}=0$,
which means that the magnetic traction vanishes at the surface of the star
(e.g., Saio \&
Gautschy 2004; Lee 2005, 2007, 2008). 
The jump conditions are given by the continuity of the function $\mbox{\boldmath $t$}$ and that of the traction 
at the two interfaces (e.g., Lee 2007, 2008). 
In addition we adopt normalization condition $T_1\left(R\right)=1$ for even modes, and $T_2\left(R\right)=1$ for odd modes at the stellar surface.

With the method of calculation we employ for the oscillations of magnetized stars, 
we find numerous solutions to the oscillation equations for a given $B_{{\rm{S}}}\equiv\mu_b/R^3$. 
Most of the solutions, however, are dependent on $j_{{\rm{max}}}$, and we have to look for the solutions 
that are independent of $j_{\rm max}$.

\section{Numerical results}

\subsection{Code test}

In order to test our numerical code, we take the Newton limit ($p/\rho c^2\to 0$, $\Lambda\to 0$, and
$\Phi\to 0$) in
the oscillation equations, and we replace the inner poloidal field with a dipole
magnetic field of Ferraro (1954) type. The Ferraro type field in
Newtonian gravity is given by 
\begin{eqnarray}
B_r=\frac{1}{r^2\sin\theta}\pdn{U}{\theta}{},\quad B_\theta=-\frac{1}{r\sin\theta}\pdn{U}{r}{},
\end{eqnarray}
where $U$ is a scalar function given by
$U=(C_1r^2+C_2r^4)\sin^2\theta\equiv F(r)\sin^2\theta$ that satisfies
the Grad-Shafranov equation for uniform density stars, and $C_1$ and $C_2$ are constants to be
determined by the boundary conditions. 
We replace $a_1(r)$ in
equations (13) and (14) with the function $F(r)$ for the field of Ferraro
type. 
This replacement makes the oscillation equations the same as those solved by Lee (2008) for
axisymmetric toroidal modes of magnetized neutron stars with a solid
crust in Newtonian gravity. The results of the axisymmetric toroidal mode
calculations for this magnetic field are the same as those in Lee
(2008), that is, we find both even and odd modes, and the eigenfrequencies
form mode sequences and the frequency in a mode sequence gradually
increases as the number of radial nodes of the eigenfunction
increases.

\subsection{General relativistic calculation}

In Figure 1, the oscillation frequencies of the toroidal modes of odd
parity of NS08 (left panels) and NS14 (right panels) are plotted versus the number of radial nodes of the eigenfunction
$T_{l^\prime_1}$ for $B_{{\rm{S}}}=10^{16}$ G (top panels) and
$B_{{\rm{S}}}=5\times 10^{15}$ G (bottom panels).
Note that no toroidal modes of even parity are found in the present calculation.
The qualitative property of the frequency spectra obtained for the
seven models in Table 1 are almost the same, although as the neutron star mass decreases, 
the mode frequencies becomes higher, 
the tendency of which is the same as that found by Lee (2008).
Except for the lowest frequency modes, the oscillation frequencies seem to make one parameter mode sequences, and
the oscillation frequency in a mode sequence $O_1$ (or $O_2$) tends to
gradually decrease as the number of radial nodes increase. 
This tendency of the oscillation frequency in a mode sequence is different from that found by Lee (2008).
In the lowest frequency group indicated by $O_0$, we find only three
modes, which all have no radial nodes of the function
$T_{l^\prime_1}$. Note that in Figure 1, the three modes in $O_0$ are
not clearly discerned because their frequencies are close to each other. 
The properties of the oscillation spectrum obtained here are not necessarily the same as those found by Lee (2008), who used the Newtonian gravity and 
the Ferraro type dipole magnetic field in the interior (see \S 4.1).
The normalized frequencies $\bar\omega\equiv\omega/\omega_0$ with $\omega_0=\sqrt{M/R^3}$ obtained in the present calculation are larger than those of Lee (2008) 
and the frequency separations between the consecutive mode sequences are wider than those of Lee
(2008). In Table 2, we summarize the eigenfrequencies $\bar{\omega}$ of
axisymmetric toroidal modes of NS08 and of NS14 for $B_{\rm
  S}=10^{16}$ G.

From Figure 1, we also find that the qualitative behavior of the frequency
spectra for $B_{{\rm{S}}}=5\times 10^{15}$ G is the same as that for
$10^{16}$ G, 
although the frequency of the modes, proportional to $B_{{\rm{S}}}$,
is smaller than that for $B_{{\rm{S}}}=10^{16}$ G.
This suggests that the modes we obtain here are essentially Alfv\'en modes.
The number of discrete toroidal modes we can find for
$B_{{\rm{S}}}=5\times 10^{15}$ G is decreased compared with the case of $10^{16}$ G, and 
as $B_{\rm S}$ becomes further weaker, it becomes more and more difficult to obtain 
well converged discrete toroidal modes.
As suggested by Lee (2008), this difficulty may be caused by the existence of a dense spectrum of core magnetic modes in the frequency range of the crustal modes for a weak field $B_{\rm S}$.

In Figure 2, we plot the eigenfunctions of the three toroidal modes in the lowest frequency group $O_0$ of NS14.  
We find that the eigenfunctions $\pmb{t}$ of the three modes look similar, that is,
they all have large amplitudes only in the outer envelope, in which the solid crust resides, and have 
negligible amplitudes in the core. 
This property is different from that of the toroidal modes belonging to the mode sequences $O_1$ and $O_2$ with higher frequencies.
As shown by Figure 3, the eigenfunctions $\pmb{t}$ of the modes in the sequences $O_1$ and $O_2$
have large amplitudes only in the fluid core and negligible amplitudes in the crust region.

\begin{figure}
\begin{center}
\resizebox{0.5\columnwidth}{!}{
\includegraphics{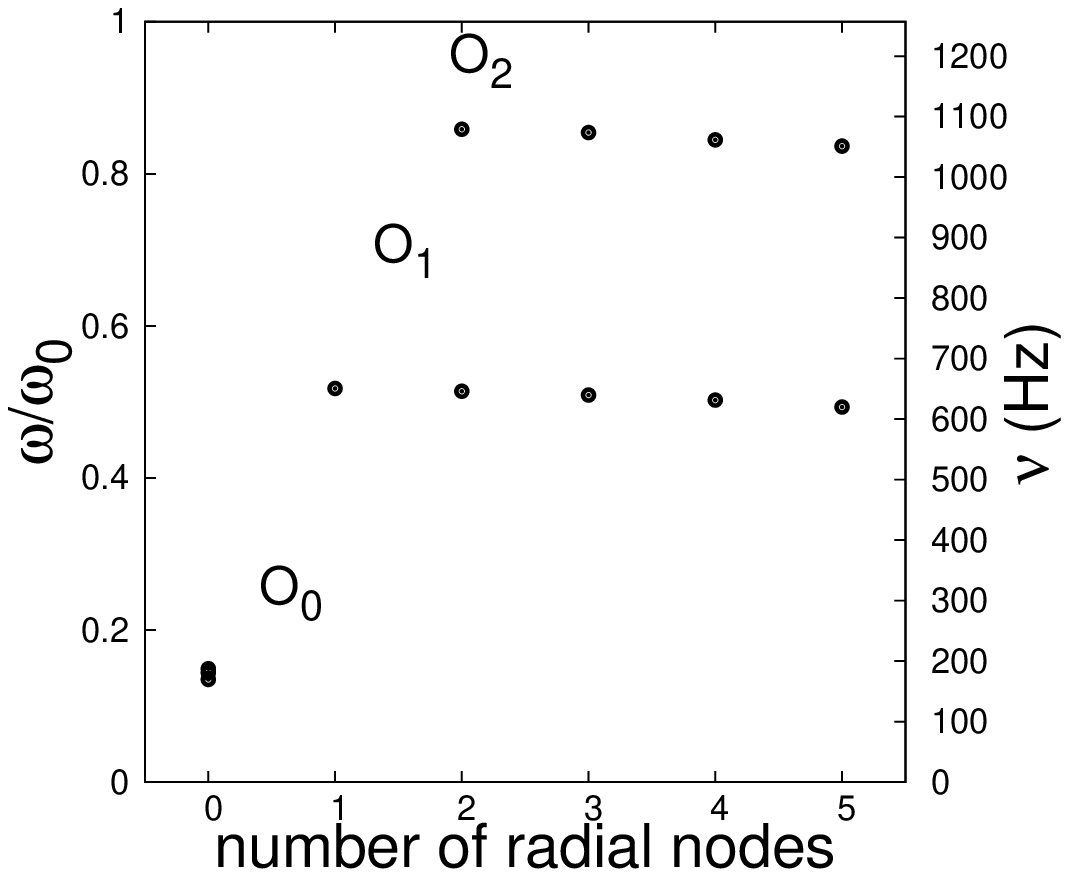}}
\hspace*{-1.5cm}
\resizebox{0.5\columnwidth}{!}{
\includegraphics{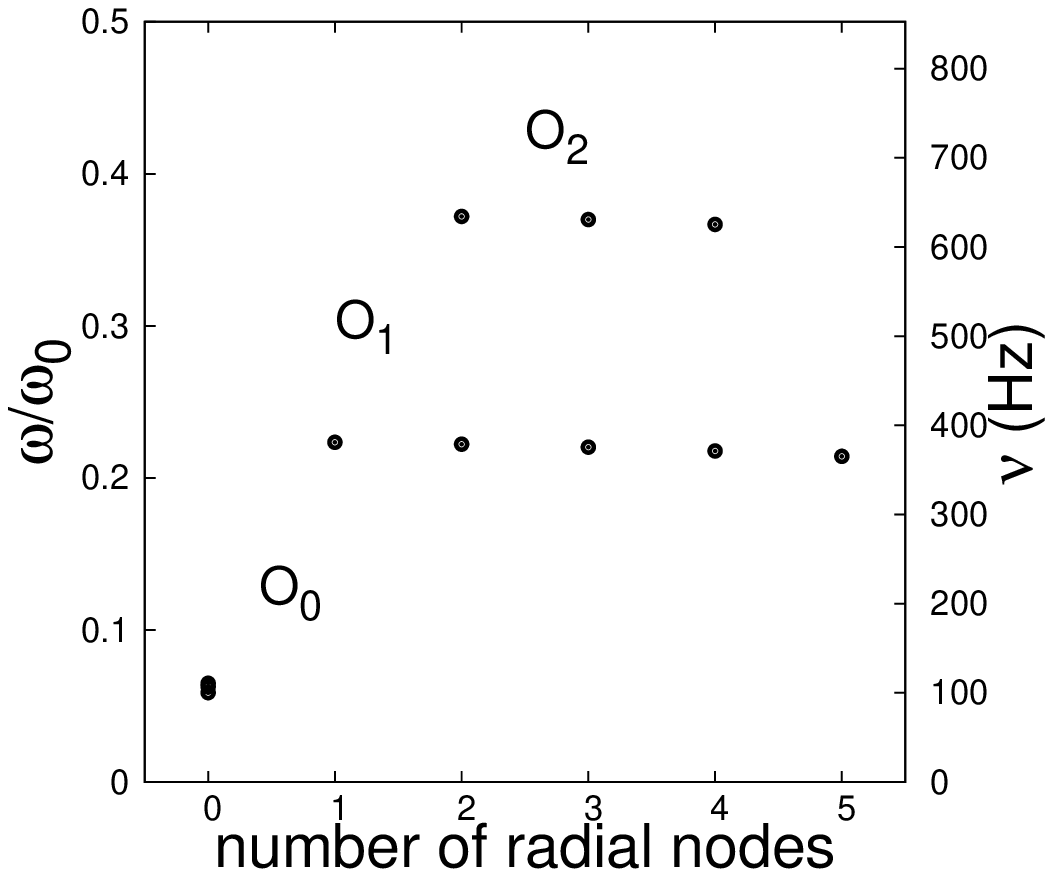}}
\resizebox{0.5\columnwidth}{!}{
\includegraphics{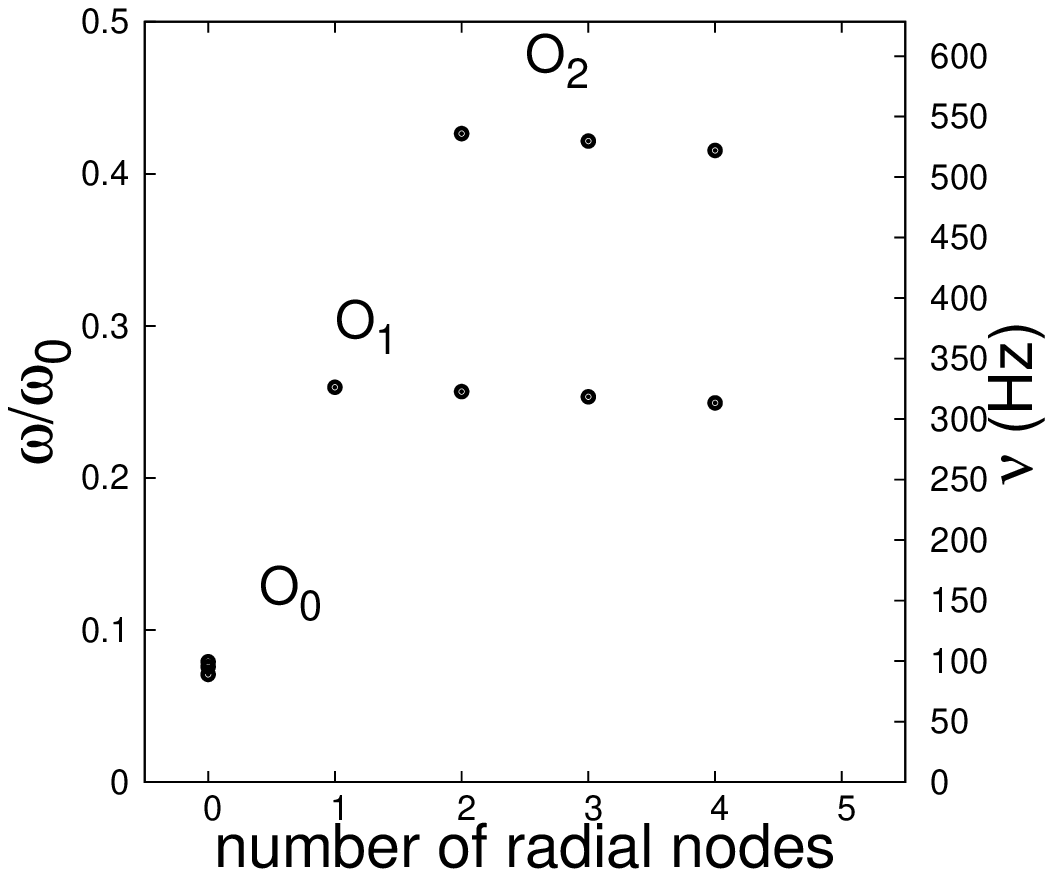}}
\hspace*{-1.5cm}
\resizebox{0.5\columnwidth}{!}{
\includegraphics{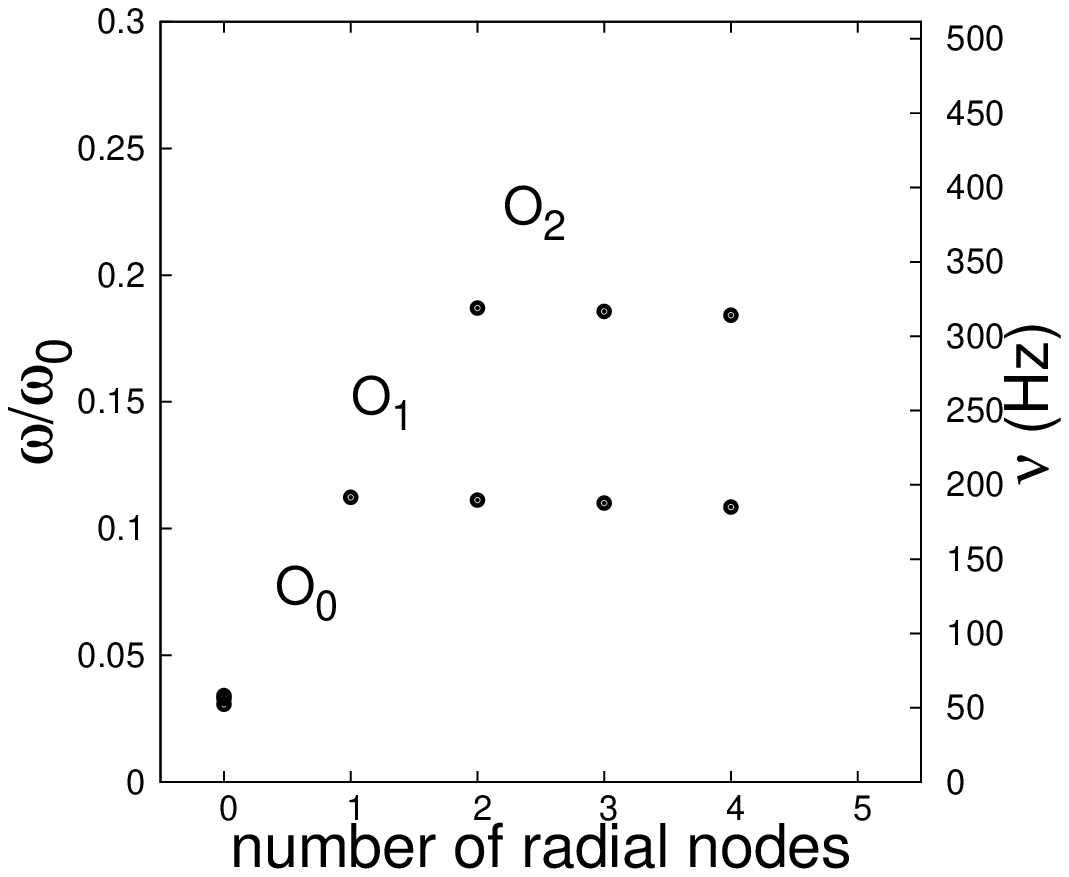}}
\end{center}
\caption{Frequencies $\omega$ of axisymmetric toroidal
  modes of NS08 (left panels) and of NS14 (right panels) versus the number of radial
  nodes of $T_{l_1'}$, where $B_{\rm S}$ is $10^{16}$ G (top panels) and $5\times
  10^{15}$ G (bottom panels). Here,
  $\nu=\omega/2\pi$ and $\omega_0=\sqrt{M/R^3}$ with $M$ and $R$ being the gravitational mass and radius of the star.}
\end{figure}

\begin{figure}
\begin{center}
\resizebox{0.39\columnwidth}{!}{
\includegraphics{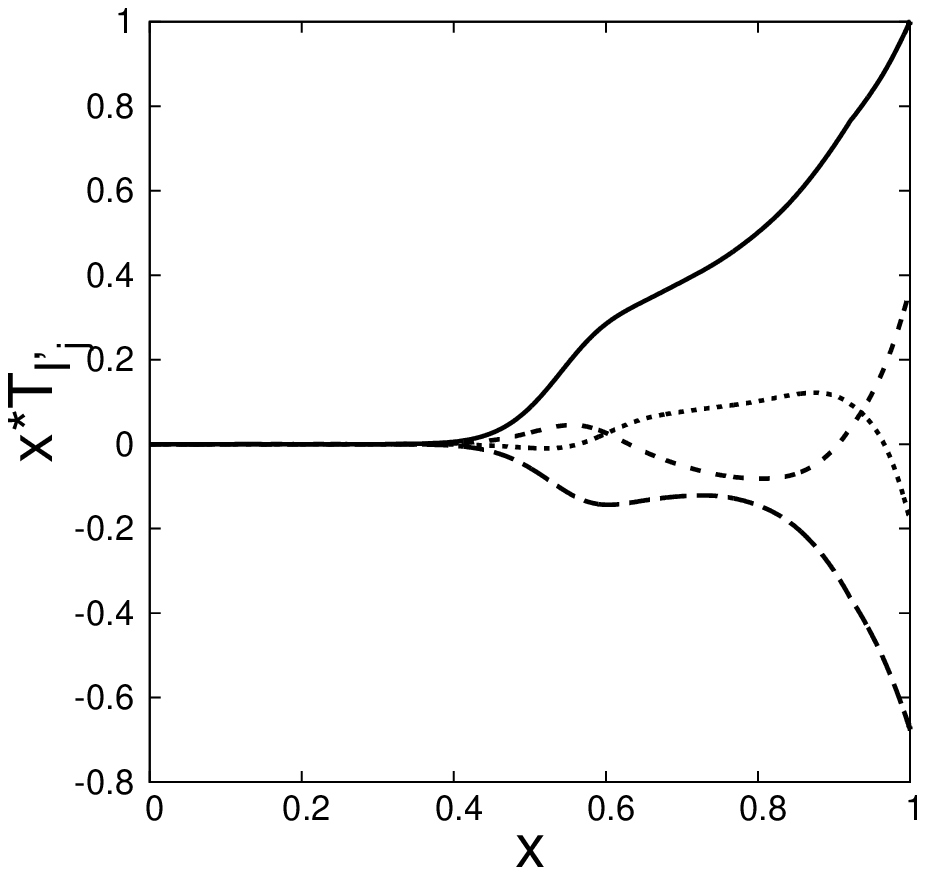}}
\hspace*{-1.75cm}
\resizebox{0.39\columnwidth}{!}{
\includegraphics{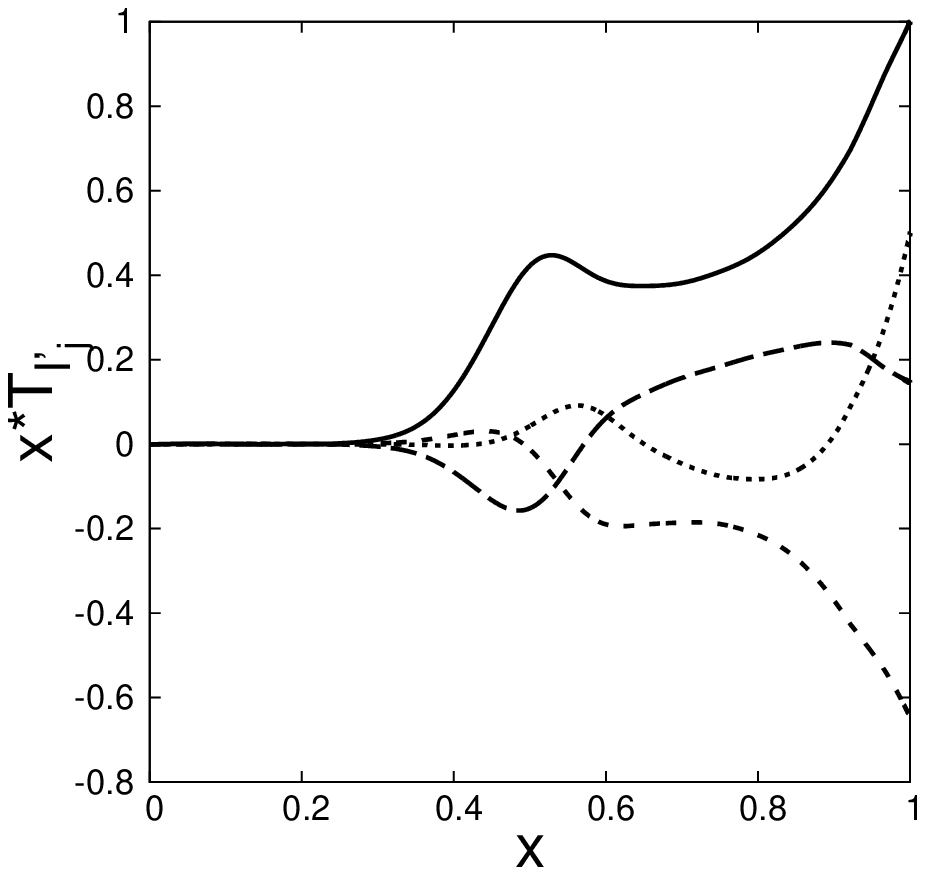}}
\hspace*{-1.75cm}
\resizebox{0.39\columnwidth}{!}{
\includegraphics{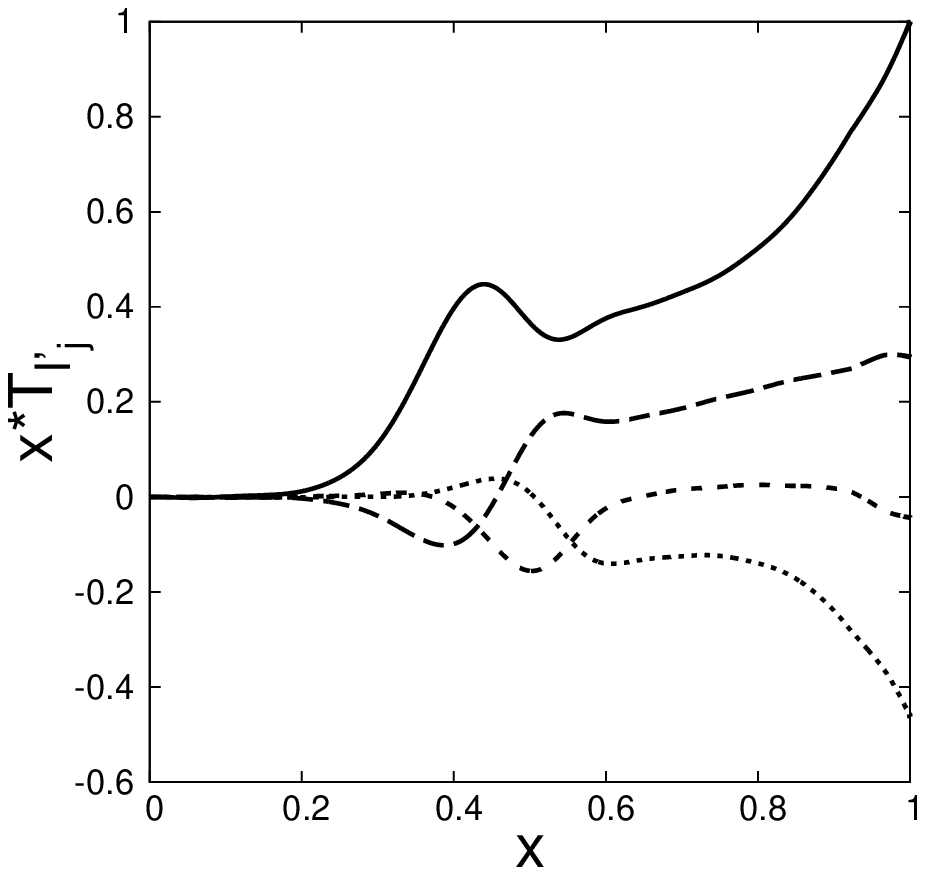}}
\end{center}
\caption{Expansion coefficients $xT_{l'}$ as a function of $x=r/R$ for the axisymmetric toroidal modes of odd parity in the $O_0$ sequence 
of the model NS14 for $B_{{\rm{S}}}=10^{16}$ G, where  the solid, long-dashed, short-dashed and dotted lines are for the expansion coefficients associated with $l^\prime_j$ from $j=1$ to 4.
Here, the frequency $\bar{\omega}\equiv \omega/\omega_0$ of the modes is, from left to right, $0.05875$, $0.06268$, and $0.06493$, respectively.
}
\end{figure}

\begin{figure}
\begin{center}
\resizebox{0.5\columnwidth}{!}{
\includegraphics{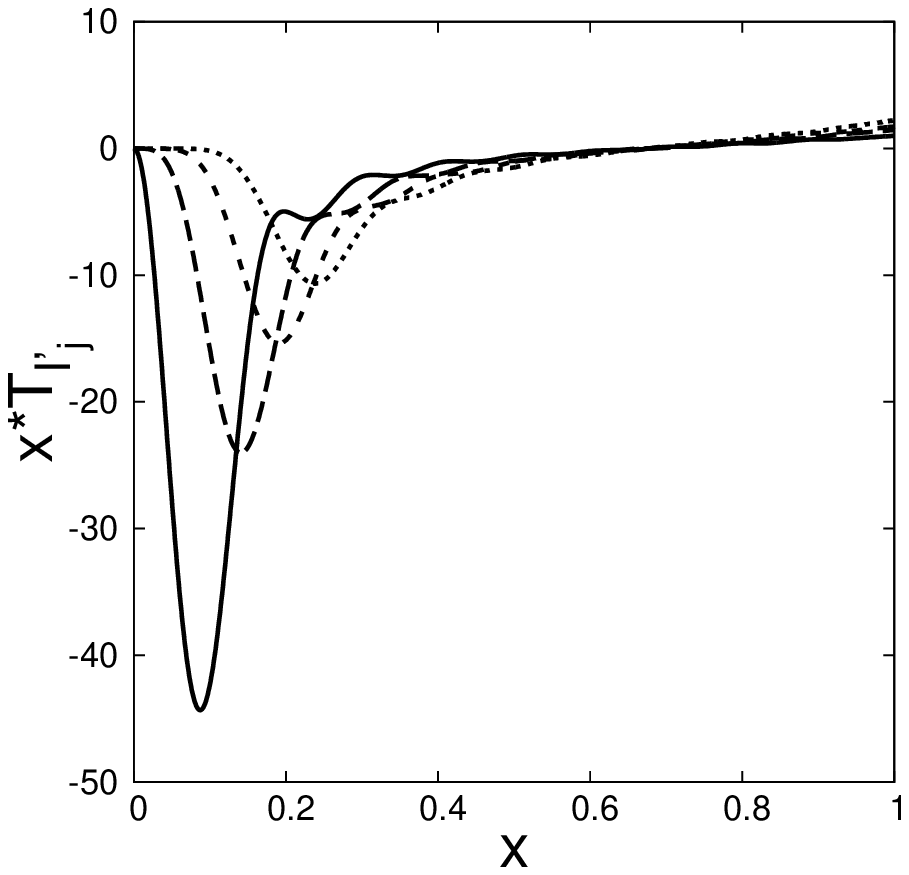}}
\hspace*{-2.5cm}
\resizebox{0.5\columnwidth}{!}{
\includegraphics{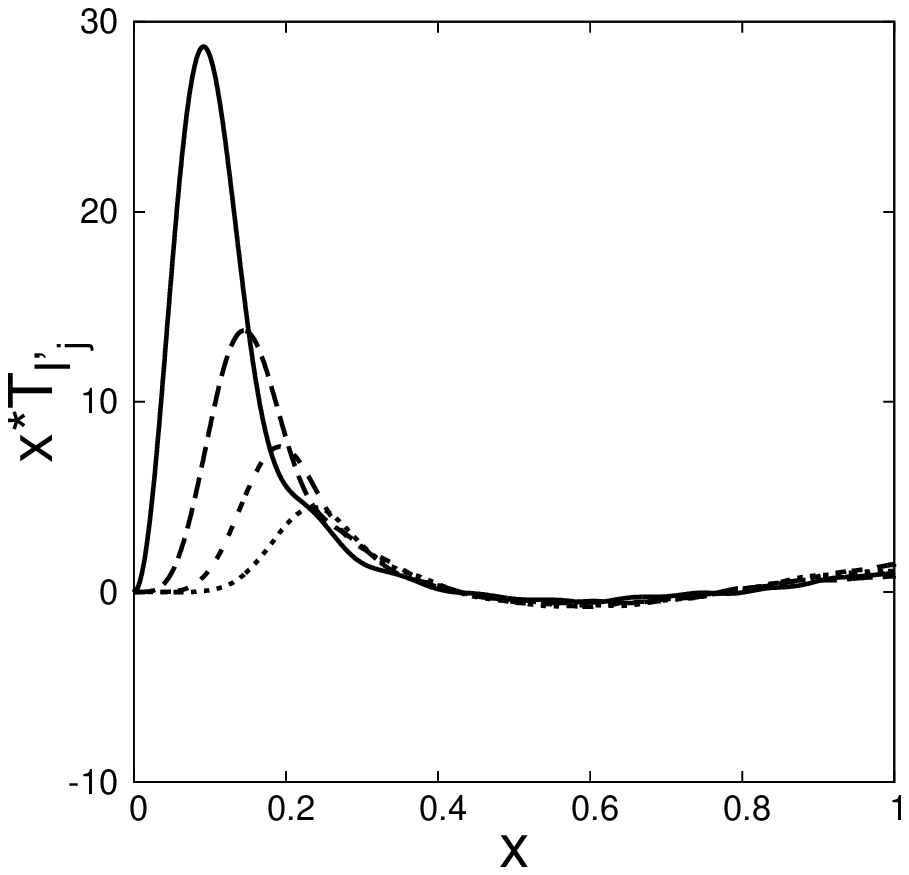}}
\end{center}
\caption{Same as Figure 2 but for the toroidal modes in the mode sequences $O_1$ (left panel) and 
$O_2$ (right panel), where the modes have the least number of radial nodes of the eigenfunction $T_{l'_1}$
in each of the mode sequences. The normalized frequency $\bar\omega$ is $0.2235$ (left panel) and 
$0.3720$ (right panel).
}
\end{figure}

\begin{table*}
\begin{center}
\caption{Eigenfrequency $\bar{\omega}$ of the axisymmetric toroidal modes for $B_{\rm S}=10^{16}$ G}
\begin{tabular}{@{}ccccccc}
\hline
 & & & NS08 & & & \\
\hline
sequence & & & number of radial nodes & & & \\
\hline
 & 0 & 1 & 2 & 3 & 4 & 5 \\
$O_0\cdots\cdots$ & 0.1351 & & & & & \\
 & 0.1440 & & & & & \\
 & 0.1492 & & & & & \\
$O_1\cdots\cdots$ & & 0.5186 & 0.5148 & 0.5106 & 0.5047 & 0.4970 \\
$O_2\cdots\cdots$ & & & 0.8593 & 0.8543 & 0.8469 & 0.8366 \\
\hline
& & & NS14 & & & \\
\hline
sequence & & & number of radial nodes & & & \\
\hline
& 0 & 1 & 2 & 3 & 4 & 5 \\
$O_0\cdots\cdots$ & 0.0588 & & & & & \\
 & 0.0627 & & & & & \\
 & 0.0649 & & & & & \\
$O_1\cdots\cdots$ & & 0.2235 & 0.2221 & 0.2203 & 0.2177 & 0.2142 \\
$O_2\cdots\cdots$ & & & 0.3720 & 0.3699 & 0.3668 & \\
\hline
\end{tabular}
\medskip
\end{center}
\end{table*}

Figure 4 shows the frequency spectrum of axisymmetric toroidal modes
for the case of $B_{{\rm{S}}}=10^{16}$ G, 
where we have treated the entire interior of the stars as a fluid, neglecting the solidity of the crust. 
We find only the toroidal modes of odd parity and no toroidal modes of even parity are found.
Many low frequency toroidal modes are found and they form the mode sequence $O_0$ such that the frequency is ordered according to the number of radial nodes of the function $T_{l^\prime_1}$.
Note that the oscillation frequencies of the toroidal modes found for the fluid star are almost the same as those for the star with a solid crust, suggesting that the magnetic field is essential to determine the modal properties of the toroidal modes.
For the fluid star, the eigenfunctions $\pmb{t}$ of the toroidal modes in the sequences $O_0$, $O_1$, and $O_2$ have large amplitudes in the core but negligible amplitudes in the envelope, the properties of which are the same as those found in Figure 3. 
In other words, only the toroidal modes in the $O_0$ sequence for the models with a solid crust have 
the distinct eigenfunctions.
It is important to note that the frequencies of the toroidal modes for the fluid star 
are almost exactly proportional to $B_{{\rm{S}}}$, and 
we can find discrete toroidal modes even for a magnetic field as weak
as $B_{{\rm{S}}}\ltsim \ 10^{14}$ G.

We may look for higher frequency toroidal modes that form
mode sequences such as $O_3$, $O_4$, and $\cdots$, but we find it 
very difficult to obtain such high frequency modes, particularly when the number of radial nodes of the eigenfunctions is increased.
Because of the difficulty,
we are not sure about whether or not we can always find toroidal normal modes of arbitrarily high frequencies.

\begin{figure}
\begin{center}
\resizebox{0.5\columnwidth}{!}{
\includegraphics{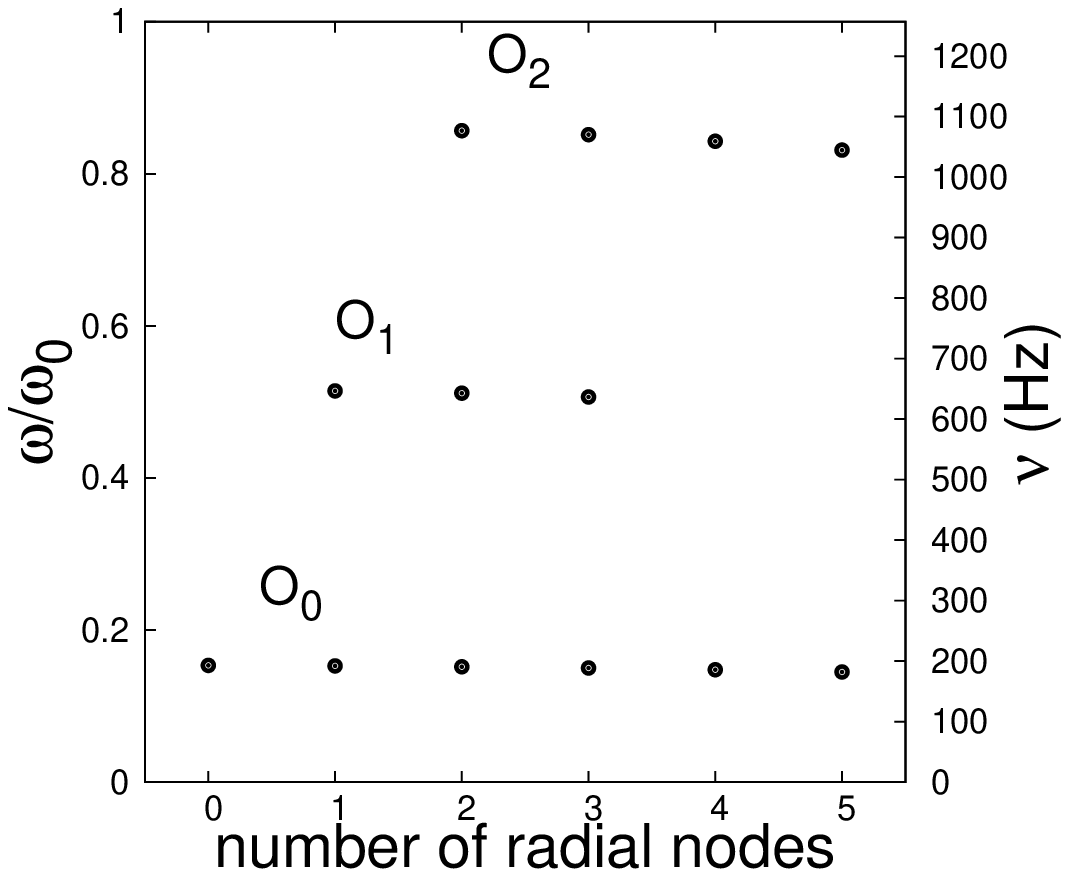}}
\hspace*{-1.5cm}
\resizebox{0.5\columnwidth}{!}{
\includegraphics{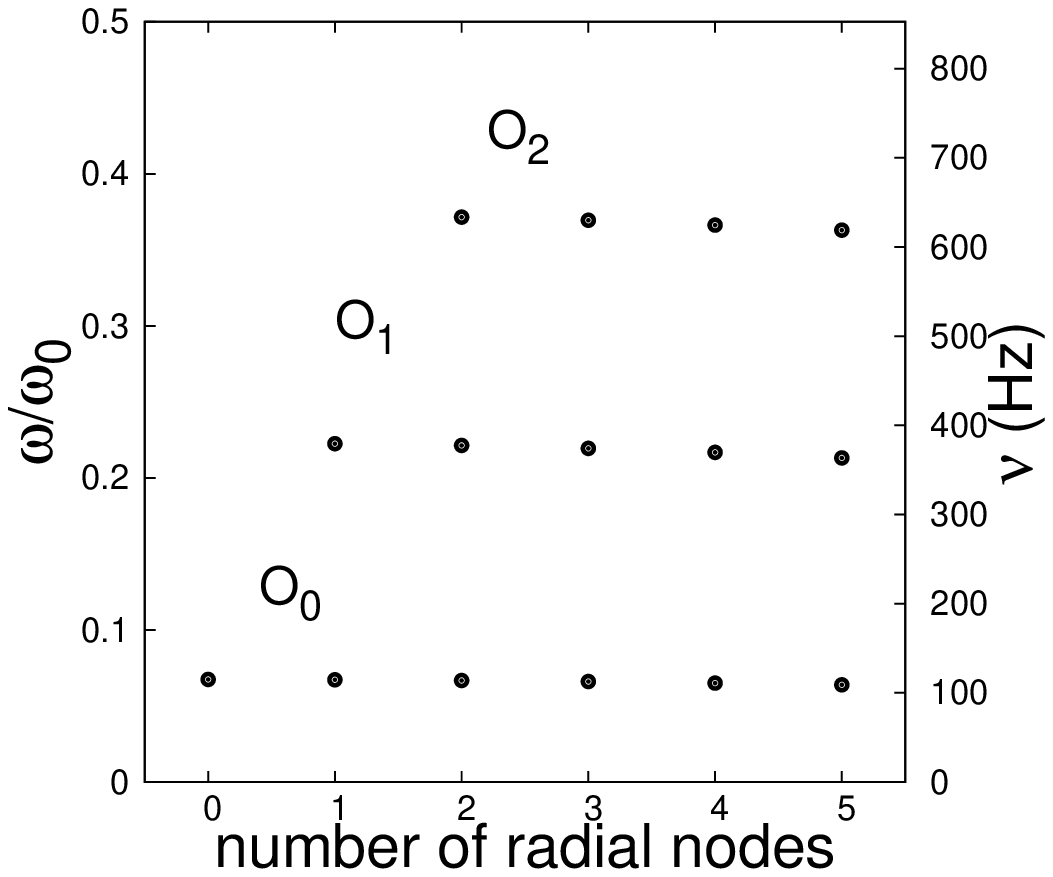}}
\end{center}
\caption{Frequencies $\omega$ of the axisymmetric toroidal modes of
  NS08 (left panel) and NS14 (right panel) versus the number of radial nodes of
  the expansion coefficient $T_{l_1'}$ for $B_{{\rm{S}}}=10^{16}$ G. 
  The modes are calculated by treating the entire
interior as a fluid.}
\end{figure}

\subsection{Newton limit}

\begin{figure}
\begin{center}
\resizebox{0.5\columnwidth}{!}{
\includegraphics{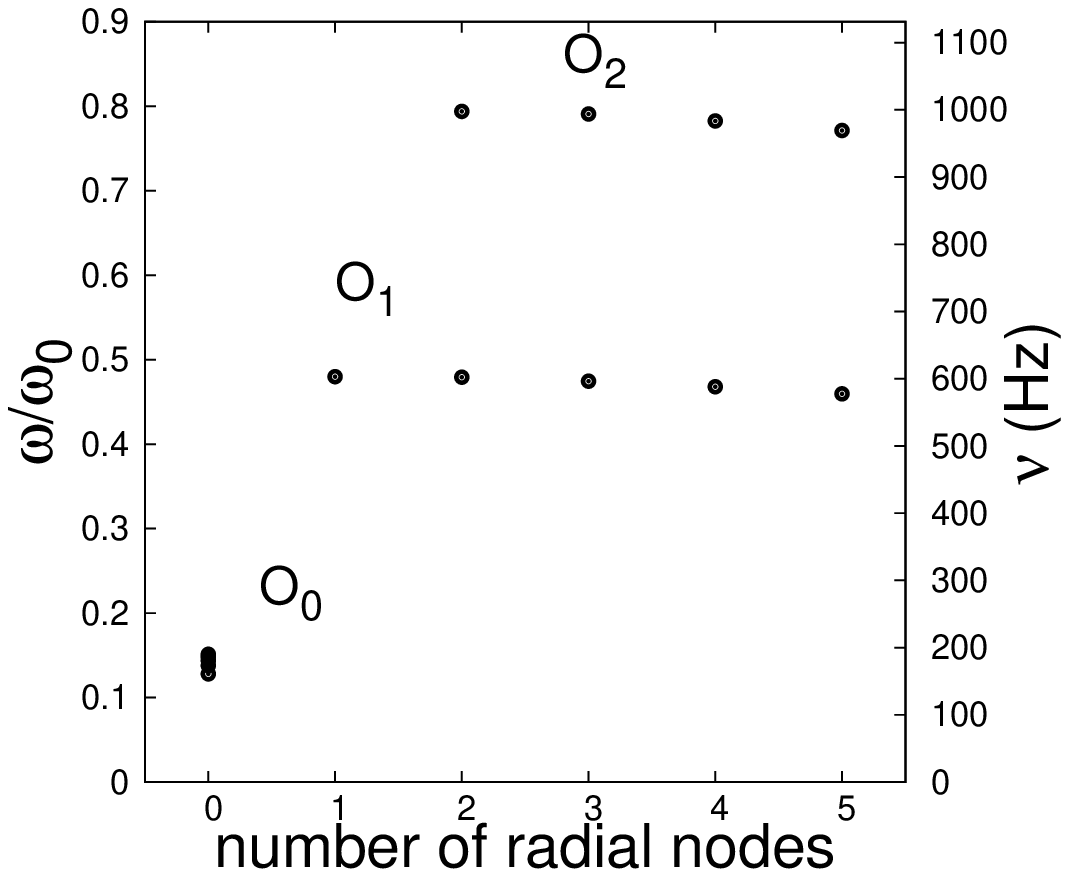}}
\hspace*{-1.5cm}
\resizebox{0.5\columnwidth}{!}{
\includegraphics{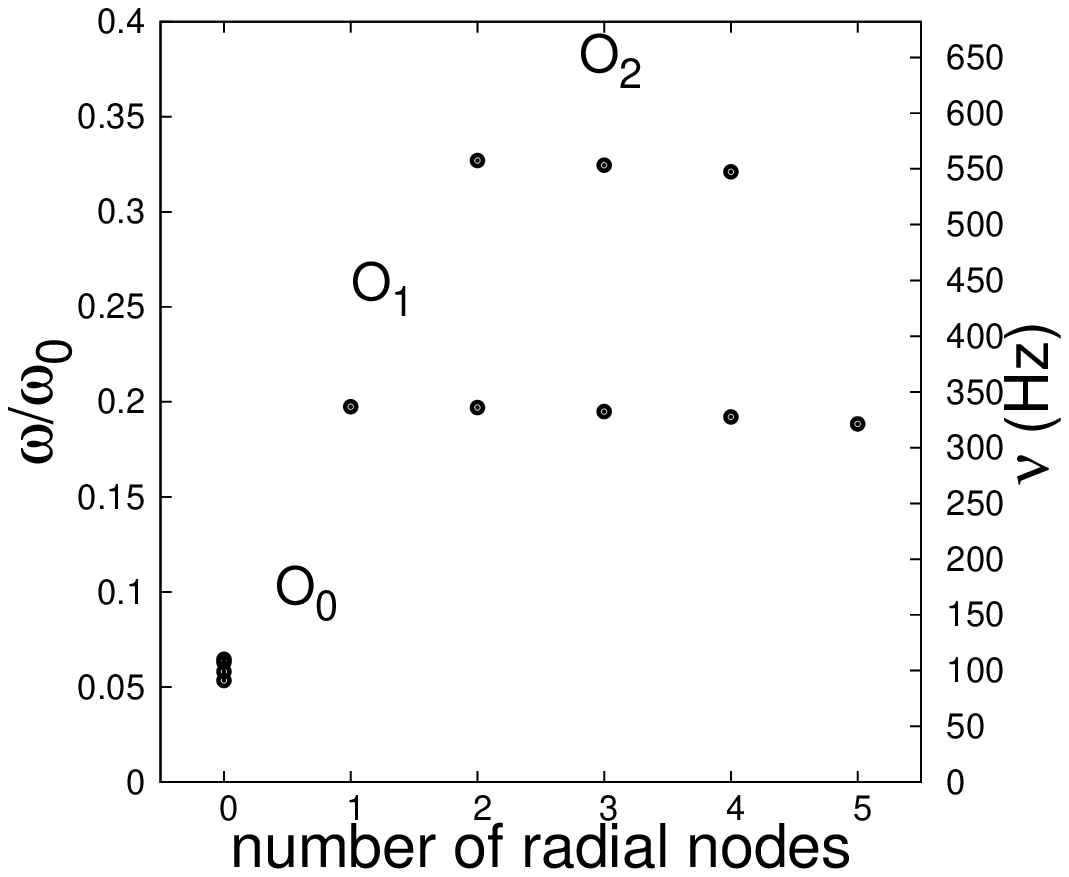}}
\resizebox{0.5\columnwidth}{!}{
\includegraphics{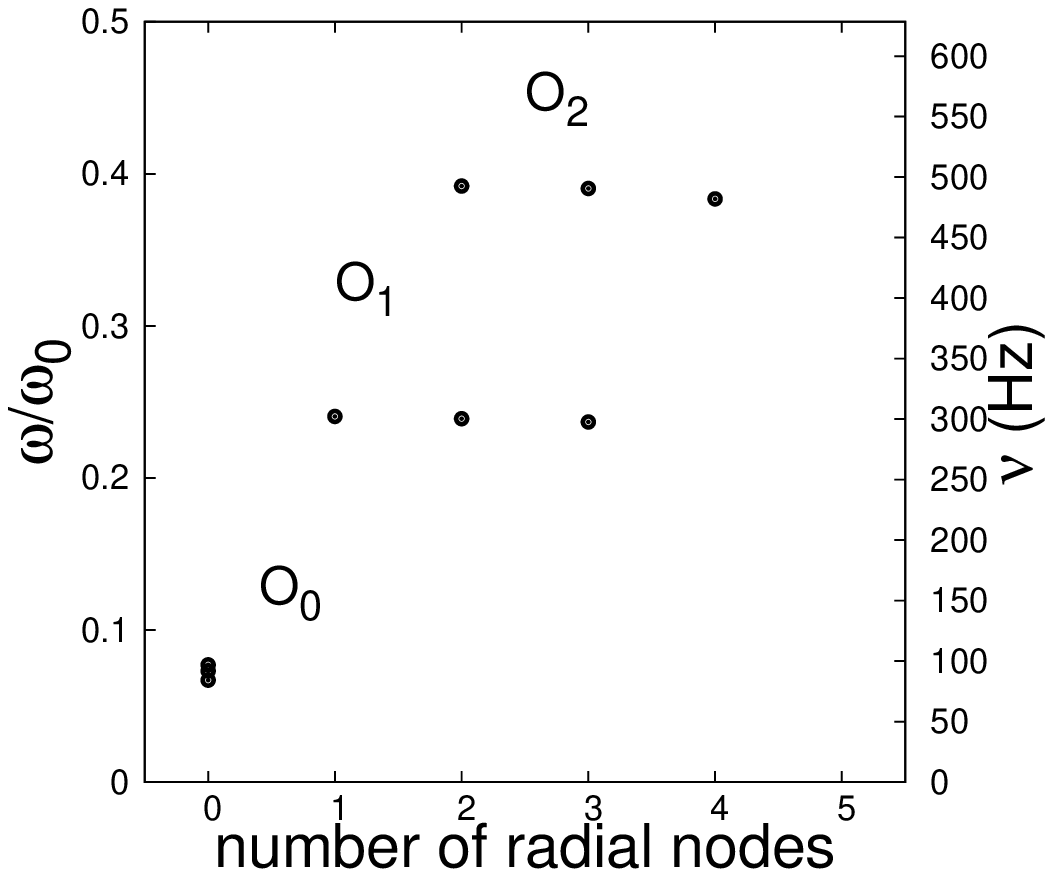}}
\hspace*{-1.5cm}
\resizebox{0.5\columnwidth}{!}{
\includegraphics{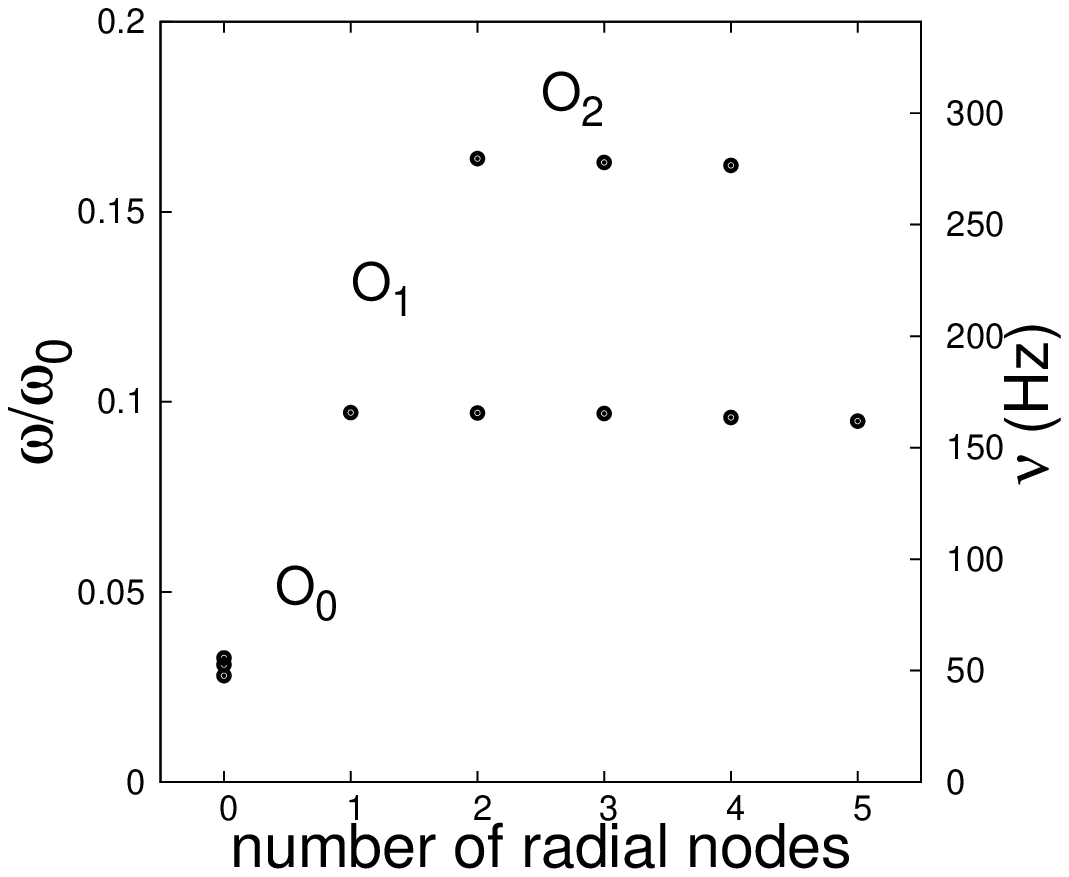}}
\end{center}
\caption{
Same as Figure 1 but for that the toroidal modes are calculated in the non-relativistic limit
of the governing equations. See the text for the detail.
}
\end{figure}

Let us consider the Newton limit of the present calculations in order to see how the general relativity affects the properties of the oscillation modes. 
The perturbation equations in the Newton limit may be obtained by assuming $p/\rho c^2\rightarrow 0$, $\Lambda\rightarrow 0$, and $\Phi\rightarrow 0$ in equations (24) and (25). 
We also take the Newton limit in the Grad-Shafranov equation for the function $a_1$. 
Figure 5 shows the results for the axisymmetric toroidal modes of NS08
(left panels) and NS14 (right panels) in the Newton limit.
Although the frequencies $\bar\omega$ are smaller, about by 10\%, than those obtained in the relativistic calculation, 
the qualitative properties of the frequency spectrum remain the same between the two calculations. 
We also confirm that the eigenfunctions are almost the same as those
found in the general relativistic calculation.

\section{Discussion}

\subsection{spatial oscillation pattern}

It is useful to examine the spatial oscillation pattern $\hat{\xi}_\phi(r,\theta)$
for the axisymmetric toroidal modes, where $\hat{\xi}_\phi(r,\theta)\equiv\
r\sin\theta\xi^\phi(r,\theta)=\sum_{j=1}^{j_{\rm max}} rT_{l'_j} (r)\partial_\theta Y_{l'_j}^0(\theta,\phi)$
and we use $j_{\rm max}=15$.
Figure 6 shows the pattern $\hat\xi_\phi(r,\theta)$ in the $(x=r\cos\theta ,y=r\sin\theta)$ plane
for the three modes of odd parity in
the mode sequence $O_0$ of the model NS14 for $B_{\rm S}=10^{16}$ G (see Figure 1), 
where 
the magnetic axis is the horizontal axis given by $y=0$ and the oscillation amplitudes are
normalized by the maximum value.
Note that these three modes have no radial nodes of $T_{l'_1}$. 
From Figure 6, we find that the oscillation pattern is antisymmetric about the equator given by $x=0$, and that the surface fluid ocean above the solid crust manifests itself as a wavy pattern along the semicircle of $(x^2+y^2)/R^2\cong1$.
We also find that the oscillation patterns have large amplitudes surrounding the region of 
closed magnetic fields in the model and the number of nodal lines, given by $\hat\xi_\phi(r,\theta)=0$, in the pattern increases with the 
oscillation frequency, which clearly provides the way of classifying the modes.
It is interesting to note that the oscillation patterns associated with the closed magnetic fields obtained by Cerda-Duran et al. (2009), who ignored the solid crust and applied the outer boundary conditions
given by $\rd\pmb{t}/\rd r=0$, have large amplitudes only in the region of the closed fields, 
although the patterns obtained in this paper have amplitudes in the regions surrounding the closed fields.

In Figure 7, we plot the pattern $\hat{\xi}_\phi(r,\theta)$ for the modes belonging to 
the sequences $O_1$ and $O_2$ in Figure 1, where only the mode with the least number of radial nodes of $T_{l'_1}$ is shown for each of the mode sequences.
From Figure 7, we find the oscillation patterns have large amplitudes in the narrow regions ($0\la y/R\la 0.3$) parallel to the magnetic axis ($y=0$).
We also find a wavy pattern along the surface fluid ocean, particularly for the mode in $O_2$.
The oscillation patterns are quite similar to those given by Cerda-Duran et al. (2009).
Note that although only odd parity modes are found in this paper, Cerda-Duran et al. (2009) found both
even and odd parity modes.

Figure 8 shows the oscillation patterns of the toroidal modes of odd parity calculated by 
treating the entire interior as a fluid, where the modes of the least number of radial nodes of $T_{l'_1}$ in each of the mode sequences $O_0$, $O_1$, and $O_2$ are shown.
The oscillation patterns of the modes in the sequences $O_1$ and $O_2$ look quite the same between the models
with and without the solid crust.
However, the patterns of the modes in the sequence $O_0$ are significantly different between the two cases.
For the model without a solid crust, we find
only the modes that have large amplitudes along the magnetic axis. 
No modes are found that have dominant amplitudes in the regions surrounding 
the closed magnetic fields.
We note that the number of nodal lines of $\hat\xi_\phi(r,\theta)=0$ in the $(x,y)$ plane increases
with the mode frequency.

It is confirmed that both the eigenfrequency
and eigenfunction of the toroidal modes do not change even if we double the number of radial mesh points
in the model, where we usually use neutron star models having about 1000 mesh points.
Note that the eigenfunctions we obtain for the toroidal modes have no indications of 
discontinuities in the derivative $\rd\pmb{t}/\rd r$.
It is also confirmed that the oscillation patterns as given in Figures 6 to 8 show
a good convergence for $j_{\rm max}\ga$ 10.

\begin{figure}
\begin{center}
\resizebox{0.7\columnwidth}{!}{
\includegraphics[angle=0]{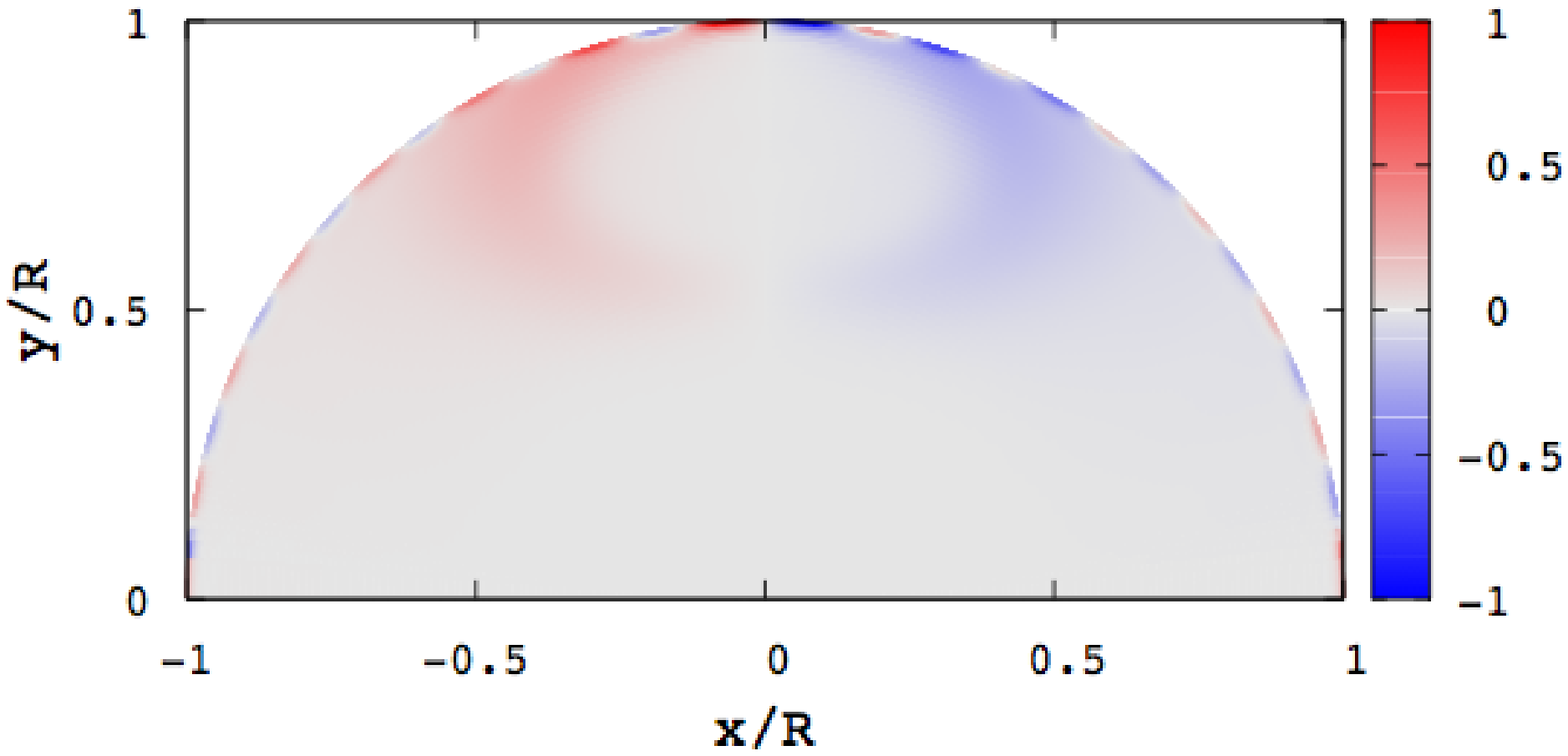}}
\hspace*{-0.35cm}
\resizebox{0.7\columnwidth}{!}{
\includegraphics[angle=0]{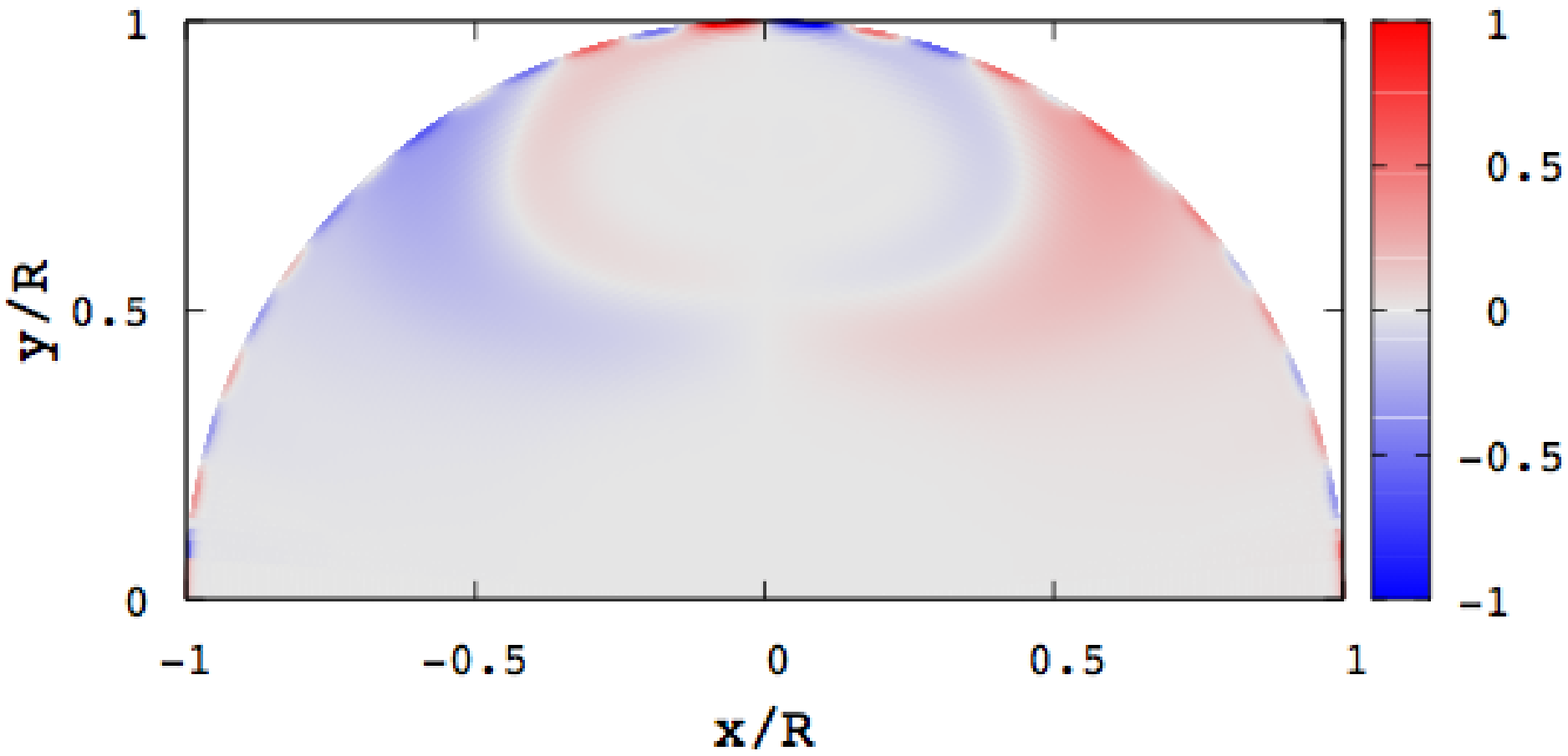}}
\hspace*{-0.35cm}
\resizebox{0.7\columnwidth}{!}{
\includegraphics[angle=0]{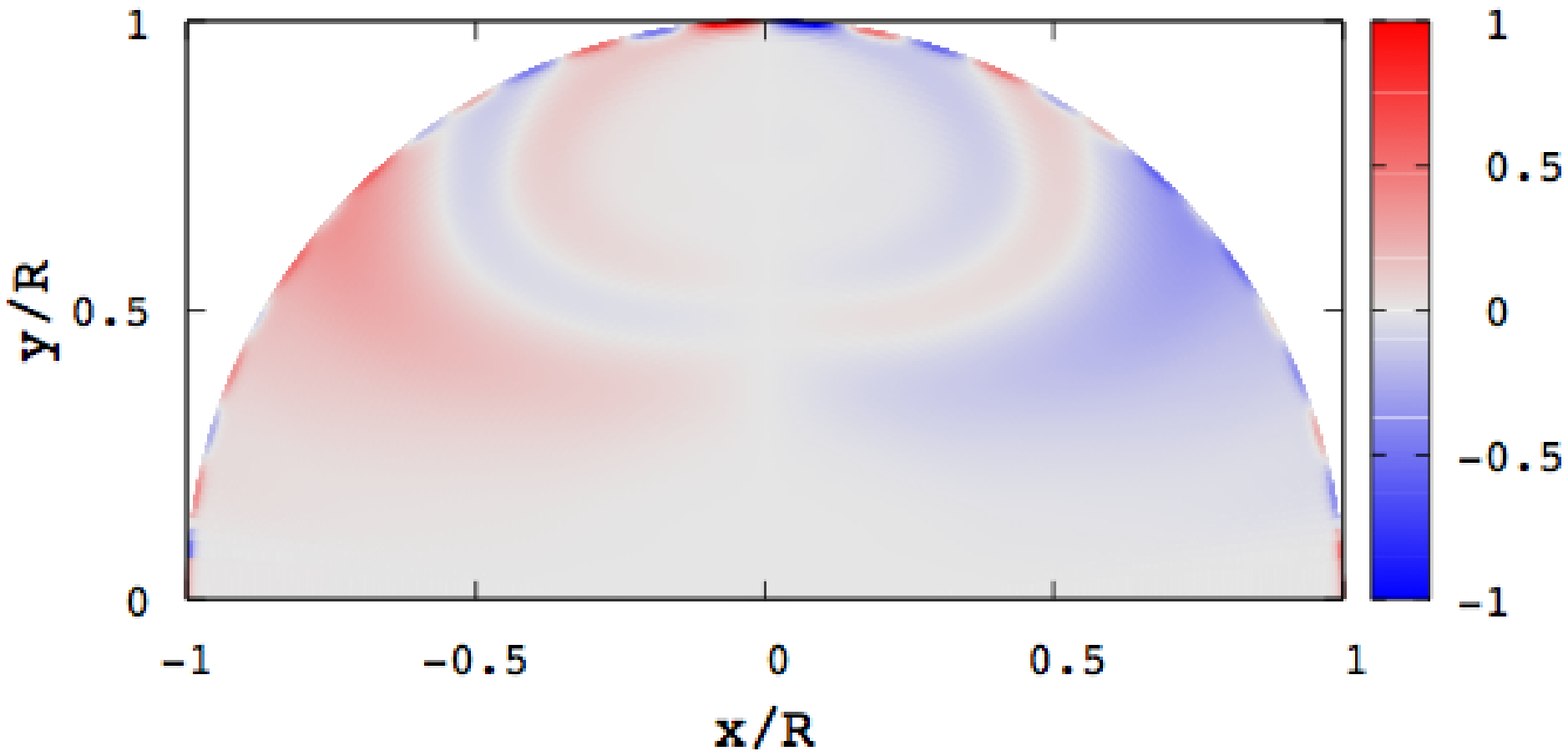}}
\caption{Spatial oscillation patterns $\hat\xi_\phi(r,\theta)$ of the axisymmetric toroidal modes of odd parity
for the model NS14 for $B_{{\rm{S}}}=10^{16}$ G, where the 
frequency $\bar\omega$ of the modes is, from top to bottom panels,
$0.05875$, $0.06268$, and $0.06493$, respectively. These
three modes belong to the lowest frequency group $O_0$ in Figure 1. }
\end{center}
\end{figure}

\begin{figure}
\begin{center}
\resizebox{0.8\columnwidth}{!}{
\includegraphics[angle=0]{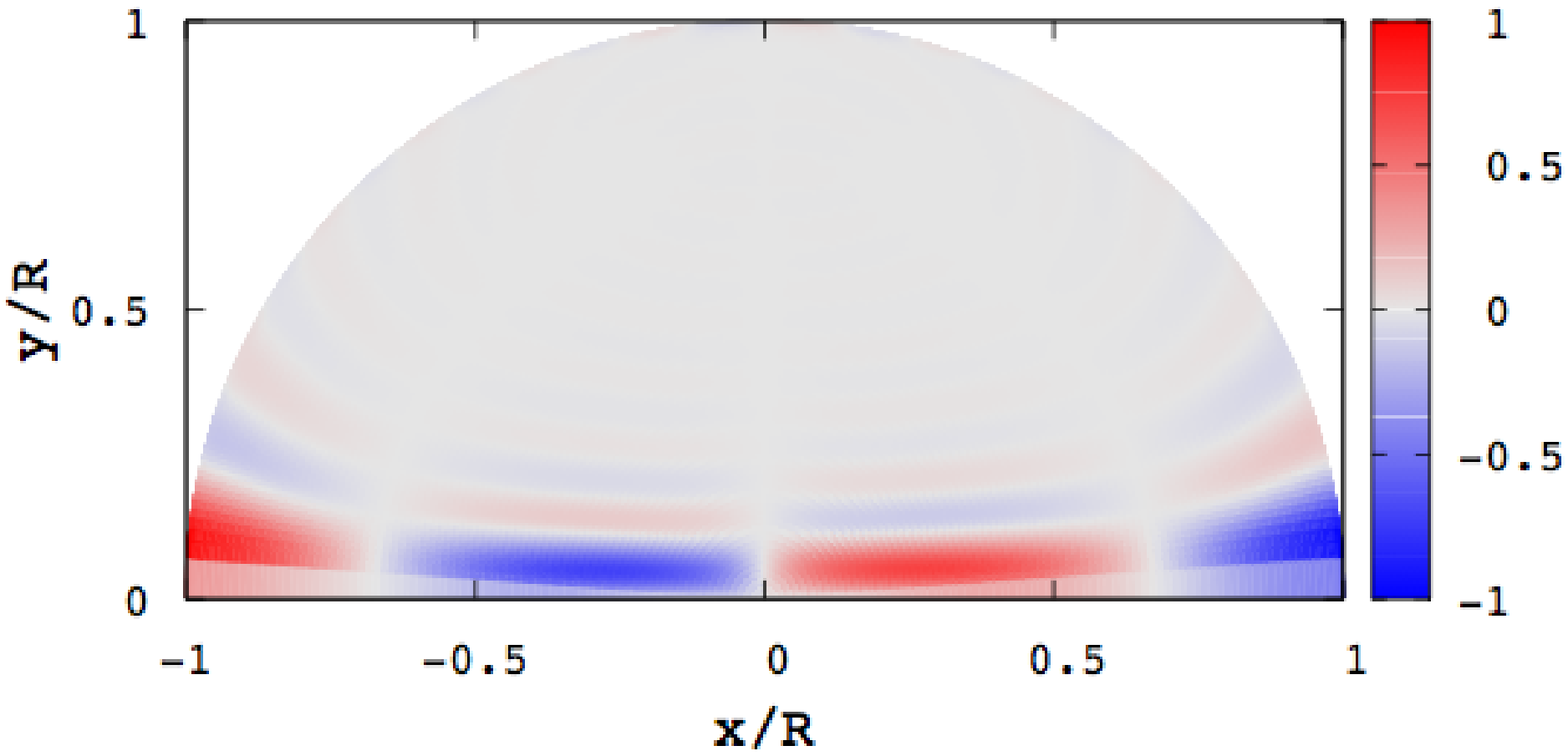}}
\hspace*{-0.38cm}
\resizebox{0.8\columnwidth}{!}{
\includegraphics[angle=0]{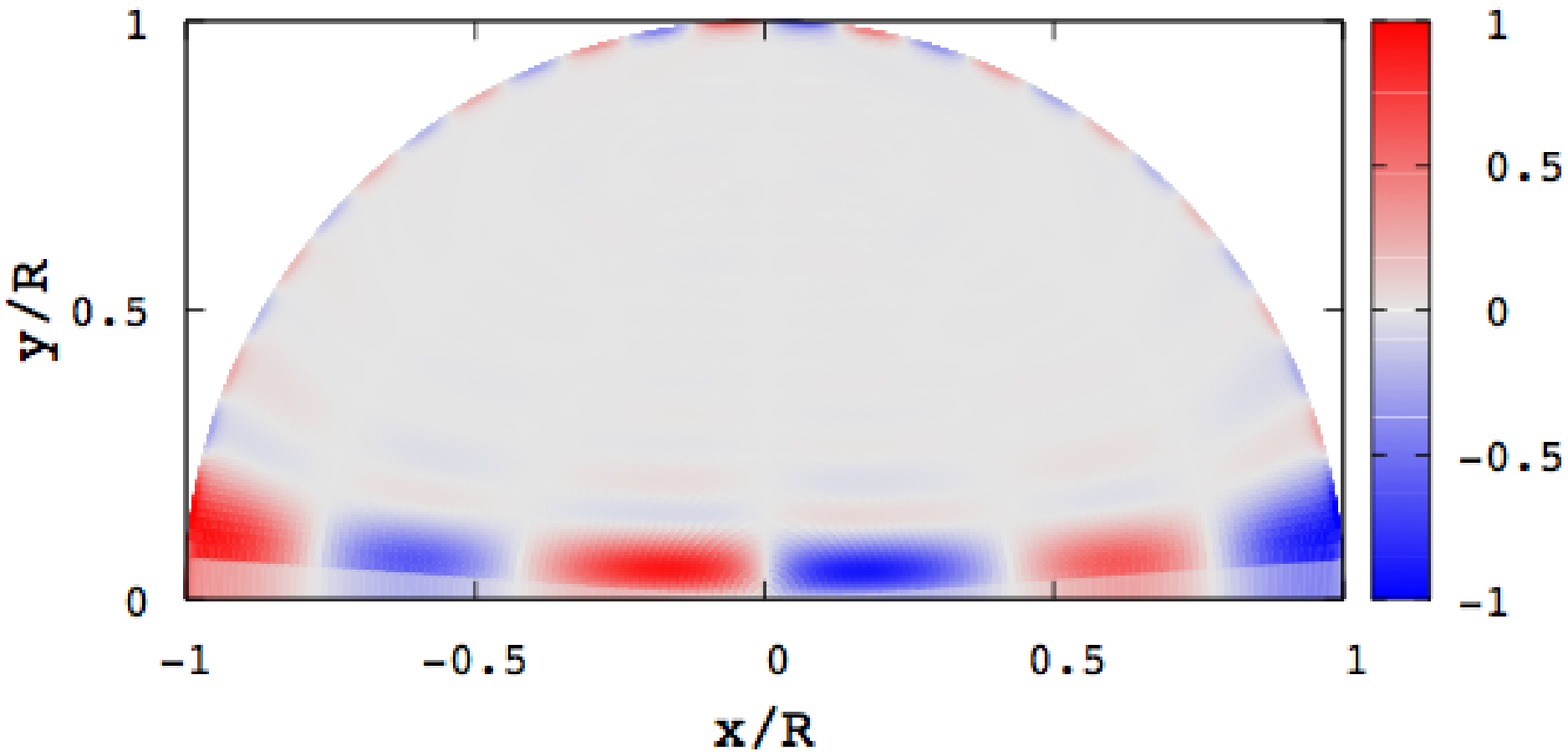}}
\caption{Same as Figure 6 but for the modes belonging to the mode sequences $O_1$ (top panel) and $O_2$ (bottom panel) in Figure 1, where 
the modes shown have the least number of radial nodes of $T_{l'_1}$ in each of the mode sequence. 
The oscillation frequency $\bar\omega$ is $0.2235$ (top panel) and
$0.3720$ (bottom panel).}
\end{center}
\end{figure}

\begin{figure}
\begin{center}
\resizebox{0.7\columnwidth}{!}{
\includegraphics[angle=0]{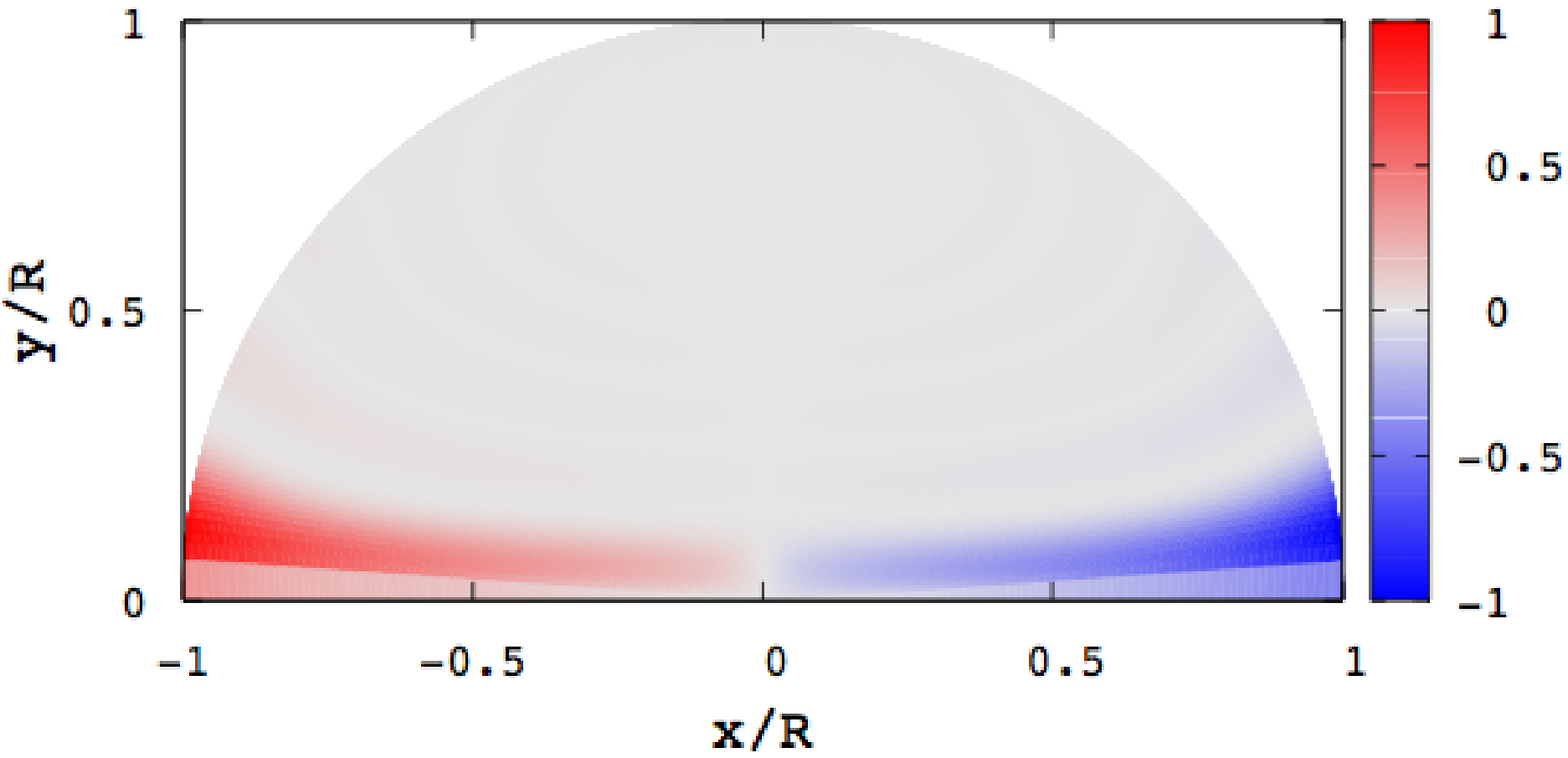}}
\hspace*{-0.35cm}
\resizebox{0.7\columnwidth}{!}{
\includegraphics[angle=0]{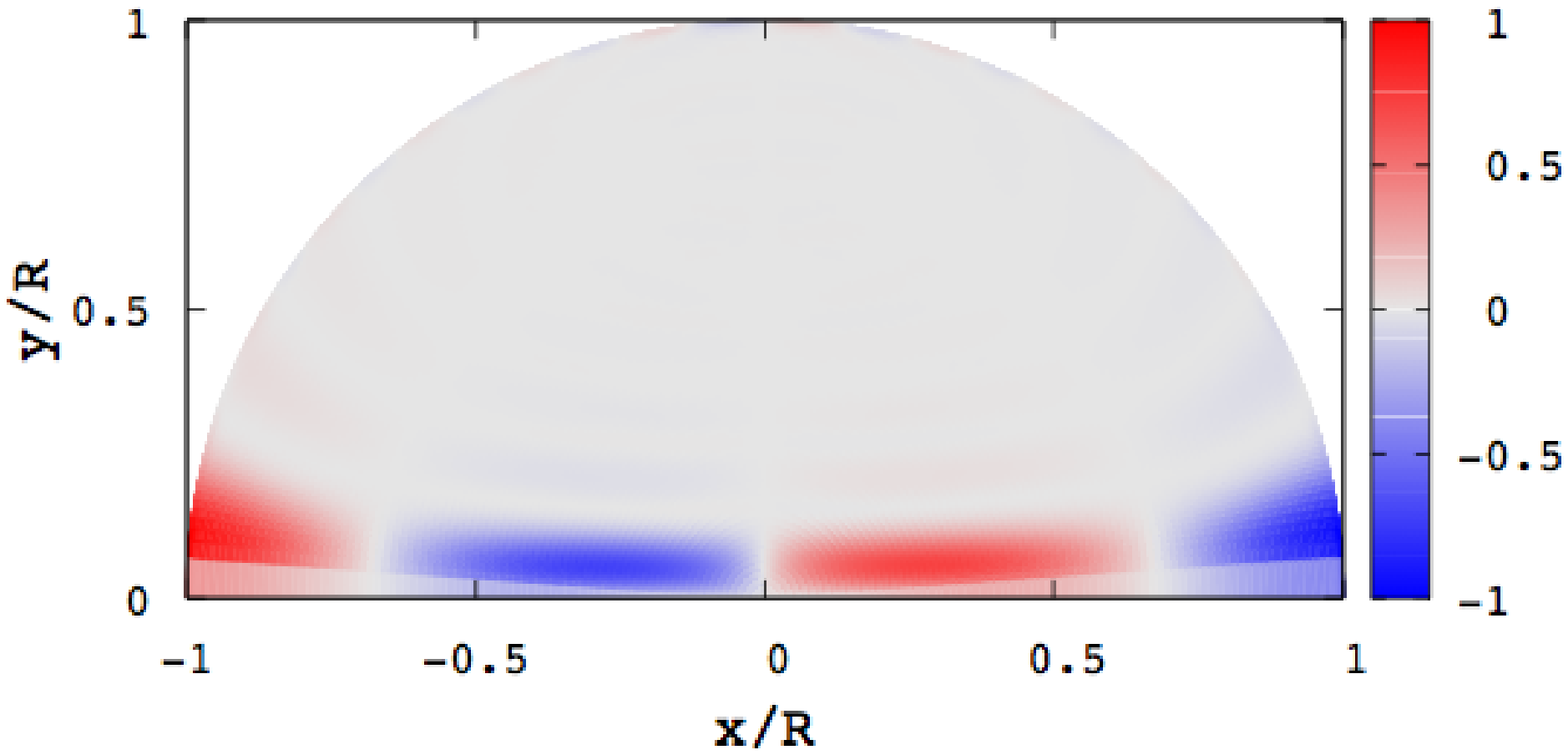}}
\hspace*{-0.35cm}
\resizebox{0.7\columnwidth}{!}{
\includegraphics[angle=0]{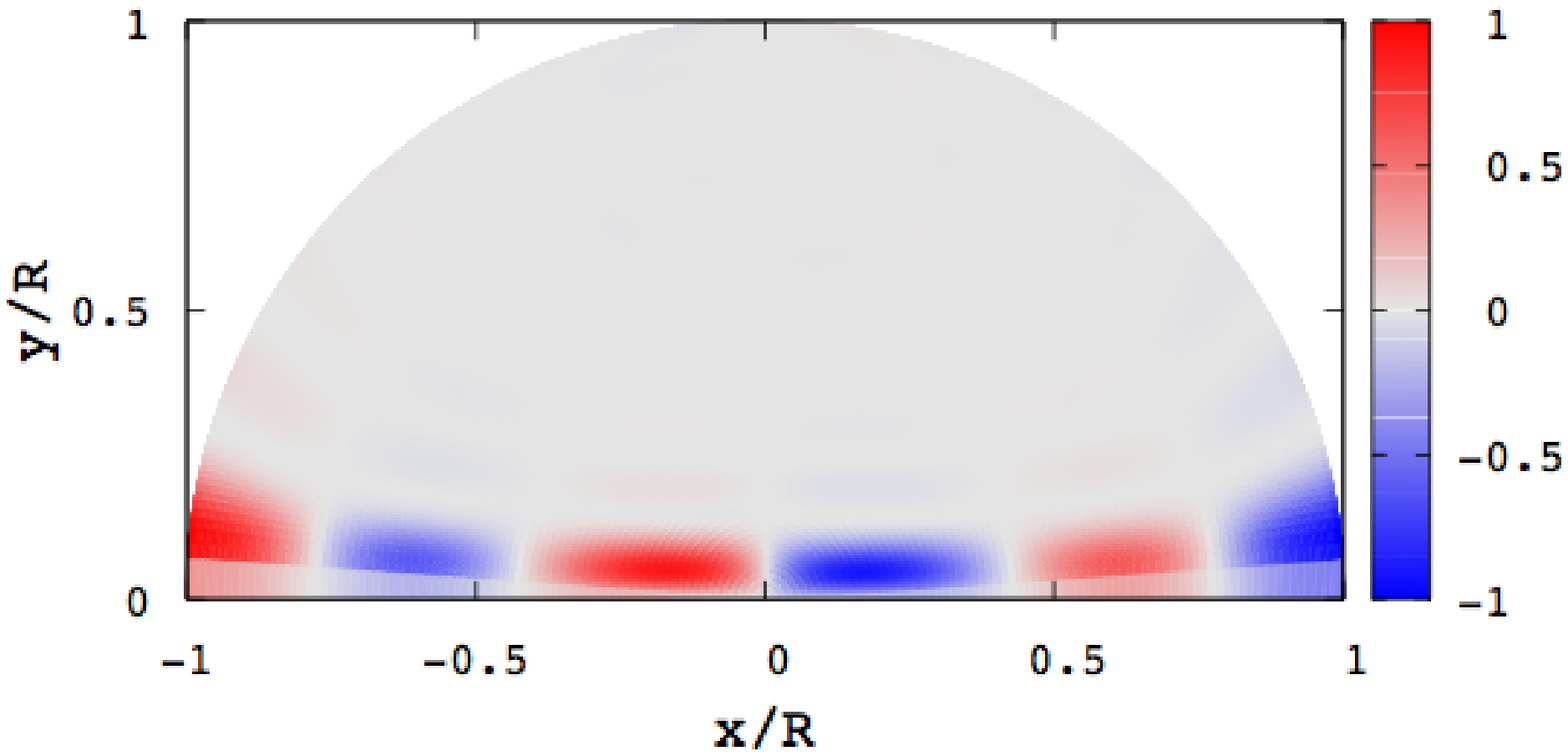}}
\caption{Same as Figure 7 but for the modes, from top to bottom panels, belonging to the mode sequences $O_0$, $O_1$, and $O_2$ in Figure 4, where 
the modes have the least number of radial nodes of $T_{l'_1}$ in each of the mode sequences.
Here, the modes are
calculated by treating the entire interior as a fluid.}
\end{center}
\end{figure}

\subsection{dependence on the compactness}

To see the dependence of the mode frequency on the compactness of the models, we plot in Figure 9
the oscillation frequency $\nu$ of three toroidal modes as a function of 
the ratio $z\equiv R/M$ for $B_{{\rm{S}}}=10^{16}$ G,
where the modes, which belong to the sequences
$O_0$, $O_1$, and $O_2$ and have the least number of nodes in each of the sequences, are calculated for the NS models tabulated in Table 1.
Applying a least-square fit to the frequency $\nu$ of the mode, for example, in the sequence $O_0$,
we obtain a fitting formula, which is in a good approximation
given by a linear function of the parameter $z$ for a given $B_{{\rm{S}}}$.
Since the toroidal mode frequency is almost exactly
proportional to $B_{{\rm{S}}}$ in the range of $B_{{\rm{S}}}$ we examined, we can write
the oscillation frequency as
\begin{equation}
\nu(z,y)\simeq \left(c_0+c_1 z\right)y,
\label{eq:31}
\end{equation}
where $y=B_{{\rm{S}}}/(10^{16} \ {{\rm{G}}})$.
In Table 3, the coefficients $c_0$ and $c_1$ are tabulated for the
first three modes in each of the mode sequences $O_0$, $O_1$, and
$O_2$. We have also carried out the same calculations by replacing the
DH EOS in the core with the APR EOS (Akmal, Pandharipande \& Ravenhall
1998), and we obtain quite similar frequency spectra of axisymmetric toroidal modes, 
and the coefficients $c_0$ and $c_1$ have also similar values to those computed for the DH EOS.

\begin{table}
\begin{center}
\caption{The coefficients $c_0$ and $c_1$ for the linear interpolation
  formulae for DH equations of state.
}
\label{symbols}
\begin{tabular}{@{}lrr}
\hline
  & $c_1$ & $c_0$\\
\hline
$O_0$ & 16.123 & 7.797 \\
      & 16.975 & 9.614 \\
      & 17.619 & 9.877 \\
$O_1$ & 62.183 & 18.543 \\
      & 62.801 & 17.843 \\
      & 63.337 & 16.988 \\
$O_2$ & 102.513 & 36.627 \\
      & 103.579 & 35.997 \\
      & 104.339 & 34.316 \\
\hline
\end{tabular}
\medskip
\end{center}
\end{table}

\begin{figure}
\begin{center}
\resizebox{0.5\columnwidth}{!}{
\includegraphics{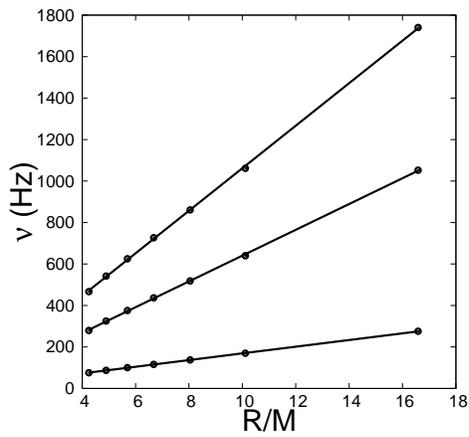}}
\end{center}
\caption{Frequencies $\nu$ of the axisymmetric toroidal modes of the neutron star models in Table 1
versus the ratio $z\equiv R/M$ for
$B_{{\rm{S}}}=10^{16}$ G, where
the modes, from bottom to top, belong to the mode sequences $O_0$, $O_1$, and $O_2$, respectively.
Note that the modes with the least number of radial nodes of $T_{l'_1}$ are plotted.
Also shown are the lines given by the fitting formula (32), where the coefficients in the formula 
are determined by applying the least-square fit to the frequencies of the mode. 
}
\end{figure}

Using the fitting formula (32), we may extrapolate the frequency of the modes
to the case of weak magnetic fields.
In Figure 10, we plot the frequency $\nu(z,y)$ as a function of $z$ for $y=0.2$ (solid lines) and $y=0.1$ (dashed lines).
For example, the QPO frequencies observationally detected for SGR 1806-20 are 
16.9, 18, 21.4, 30, 36.8, 59.0, 61.3, 92.5, 116.3, 150, and 626 Hz
(Israel et al 2005; Strohmayer \& Watts 2006; Hambaryan et al 2011).
If we assume $y\sim 0.2$ and $z\sim 5$, the frequencies $\sim 20$, $\sim 60$, and $\sim 110$ Hz may be
reproduced by the formula, 
but the frequencies around $\sim30$ Hz and $\sim 92$ Hz may not.
On the other hand, if we assume $y\sim 0.1$ and $z\sim 15$, the formula can explain
the frequencies $\sim 30$, $\sim 90$, and $\sim 150$, but the QPOs of
$\sim 20$ Hz and $\sim 60$ Hz cannot be fitted by the extrapolation formula.
This suggests that it is difficult to
explain all the observed QPO frequencies in terms of the toroidal normal modes alone, 
computed in this paper.
For SGR 1900+14, the QPO frequencies are $\sim28$, 53.5, 84, and 155 Hz (Strohmayer \& Watts 2005), and
these frequencies, except for 53.5 Hz, may be fitted by the formula (32)
if we assume $y\sim 0.1$ and $z\sim 14$, suggesting SGR 1900+14 is a low mass neutron star
having $\sim 0.7M_\odot$\footnote{Using the EOS APR, Colaiuda \& Kokkotas (2011)
have estimated a mass $M=1.4M_{\odot}$, a radius $R=11.57$ km and a
magnetic field $B=4.25\times 10^{15}$ G for SGR
1900+14. They have identified the three QPO frequencies 28, 53, 84 Hz
as discrete Alfv\'en modes.}.
However, it is also difficult to expect that all the detected QPOs are interpreted 
in terms of axisymmetric toroidal normal modes alone.
It is therefore fair to say that the estimations for the values of $y$ and $z$ given above
should be regarded as inconclusive.

To explain the low frequency QPOs having $\nu\la 20$ Hz in terms of
the toroidal normal modes, we probably need to assume a weak field
$B_{\rm S}$ of order of $10^{15}$ G.
But, if this is the case, we have a difficulty in explaining high
frequency QPOs as high as $\nu\sim 626$ Hz
using the toroidal normal modes.

\begin{figure}
\begin{center}
\resizebox{0.5\columnwidth}{!}{
\includegraphics{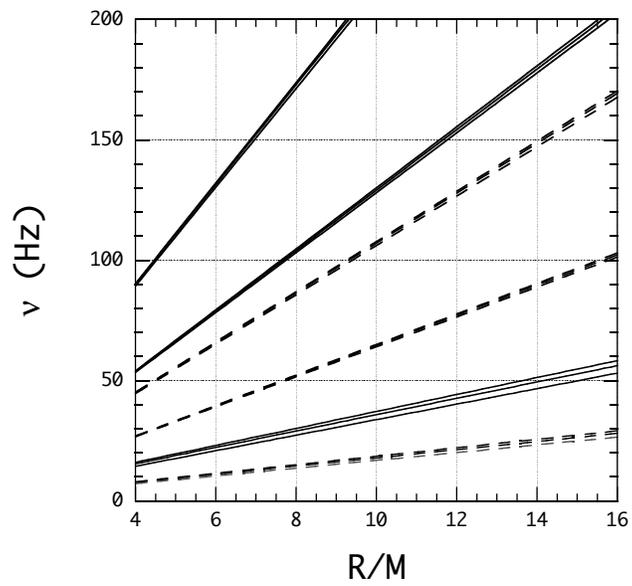}}
\end{center}
\caption{Frequencies $\nu$ of the first three toroidal modes in each of the mode sequences $O_0$, $O_1$, and $O_2$
versus the ratio $z\equiv R/M$ for $y\equiv B_{{\rm{S}}}/(10^{16} \ {\rm G})=0.2$ (solid lines) and $y=0.1$ (dashed lines).
}
\end{figure}

\section{Conclusion}

In this paper we have calculated axisymmetric toroidal modes of
magnetized relativistic neutron stars with a solid crust, where
the interior of the star is assumed to be threaded by a poloidal magnetic field that is continuous at the surface with an outside dipole field. 
We have obtained axisymmetric toroidal modes of odd parity as normal modes, 
just as computed by Lee (2008) in the Newtonian formulation.
However, no toroidal modes of even parity are found in the 
present calculation, and the reason for this is not yet well understood.
The frequency of the modes is proportional to the magnetic field strength $B_{{\rm{S}}}=\mu_b/R^3$ measured at the surface, 
suggesting that the modes we obtained are discrete Alfv\'en modes. 
We have also calculated axisymmetric toroidal modes in the Newton limit, and we obtained almost the same frequency
spectra as those in the general relativistic calculation.
In this Newton limit, assuming the Ferraro (1954) type poloidal magnetic field, 
we can reproduce the results by Lee (2008).
This suggests that
the difference between the frequency spectra obtained in this paper and by Lee (2008) may be attributable to the difference in 
the configuration of the poloidal magnetic fields assumed.

Assuming strong magnetic fields ranging from $B_{\rm S}\sim
5\times 10^{15}$ G to $10^{16}$ G,
we calculate axisymmetric toroidal modes for neutron stars of different masses.
From the frequency spectra thus computed,
we find that the frequency is almost exactly proportional to $B_{{\rm{S}}}$ and is well represented by a linear function of $R/M$
for a given value of $B_{{\rm{S}}}$.
This makes it possible to derive a fitting formula for
the frequency of the toroidal modes as a function of $R/M$ and $B_{{\rm{S}}}$.
Using thus derived fitting formula,  
we have tried to determine values of the parameters $R/M$ and $B_{{\rm{S}}}$ such that the toroidal modes can explain the detected QPOs.
Although the toroidal mode frequencies for $B_{{\rm{S}}}\sim 10^{15}$ G are in the observed QPO frequency range, 
we find it difficult to reproduce all the
QPO frequencies in terms of the toroidal normal modes alone. 
This difficulty may suggest that besides the toroidal normal modes we need  
different classes of oscillation modes for the QPOs.

As a test of the numerical method we use, taking the expansion
length $j_{\rm max}=1$ and treating the entire interior as a fluid, we calculated toroidal oscillations and
found discrete magnetic modes similar to those obtained by Sotani et al. (2006). 
The toroidal modes of odd parity in our calculation correspond to the case of $l=2$ in Sotani et al. (2006). Comparing the frequencies for the model APR+DH14 in Sotani et al (2006) with those for our NS14 model, we found that the frequencies in the two models agree.

In this paper, we reported our finding of discrete toroidal modes of magnetized stars
with a poloidal field.
This finding seems inconsistent with the results obtained by time-dependent numerical simulations, when
most of the simulations suggest only the existence of quasi-periodic oscillations (QPOs) associated with the continua in the frequency spectrum (e.g., Cerd\'a-Dur\'an et
al. 2009; Gabler et al. 2011, 2012, 2013).
However, none of the time-dependent numerical simulations could be a mathematical proof of the non-existence of discrete toroidal modes in the magnetized star.
We may expect that QPOs associated with continuous frequency spectra and discrete toroidal modes
could coexist (e.g.,  Goedbloed \& Poedts 2004).
Since our finding is based on the numerical method of a finite precision, we need a mathematically more rigorous proof concerning the existence of
discrete toroidal modes in the magnetized stars with a poloidal field.

The present analysis is a part of our study of the oscillations of
magnetized relativistic neutron stars. 
Even for a pure poloidal magnetic field, it is difficult to determine frequency spectra of non-axisymmetric toroidal modes
and those of axisymmetric and non-axisymmetric spheroidal modes.
We note that the stability of a magnetic field configuration is another difficult problem.
It is well known that a purely poloidal magnetic field is
unstable and the energy of the field is dissipated quickly, that is, for several ten milliseconds
(e.g., Lasky et al. 2011; Ciolfi \& Rezzolla 2012), although
stellar rotation may weaken 
the instability of a poloidal magnetic field as suggested by Lander \& Jones (2010). 
On the other hand, Kiuchi \& Yoshida (2008) has
calculated equilibrium configurations of rotating relativistic stars
having a purely toroidal magnetic field, expecting that some
neutron stars have a magnetic field whose toroidal components are much stronger than poloidal
ones. 
But, it was also suggested that purely toroidal magnetic
fields are unstable (e.g., Goosens 1979). 
It is thus anticipated that a mixed poloidal and toroidal magnetic field configuration
such as a twisted-torus magnetic field (e.g.,
Braithwaite \& Spruit 2004; Yoshida, Yoshida \& Eriguchi 2006; Ciolfi et al. 2009) can be
stable, that is, such a magnetic field configuration can last stably for a long time. 
If this is the case, it will be important to investigate the oscillation
modes of stars threaded by both toroidal and poloidal magnetic fields.
As suggested by Colaiuda \& Kokkotas (2012), however, the presence of a toroidal field component can significantly change the properties of the oscillation
modes of magnetized neutron stars. 
In the presence of both poloidal and toroidal field components, toroidal and spheroidal modes 
are coupled, which inevitably makes the frequency spectra complex. 

As an important physical property inherent to cold neutron stars, we need to consider
the effects of superfluidity and superconductivity of neutrons and protons on the oscillation
modes.
It is believed that neutrons become a superfluid both in the inner crust and in the fluid core while protons
can be superconducting in the core. 
For example, if the fluid core is a type I
superconductor, magnetic fields will be expelled from the core region, because of Meissner effect, and
hence confined to the solid crust (e.g., Colaiuda et al 2008; Sotani et al 2008). 
In this case, we only have to consider oscillations of a magnetized crust so long as toroidal modes are concerned, 
and we have toroidal crust modes modified by a magnetic field.
Note that spheroidal oscillations can be propagative both in the magnetic crust and in the non-magnetic fluid core.
However, a recent analysis of the spectrum of timing noise for SGR
1806-20 and SGR 1900+14 has suggested that the core region
is a type II superconductor (Arras, Cumming \& Thompson 2004). 
If this is the case, the fluid core can be threaded by a magnetic field and hence
the frequency spectra of toroidal modes will be affected by the
superconductivity in the core
(e.g., Colaiuda et al. 2008; Sotani et al. 2008).

\appendix
\section[]{Oscillation equations for relativistic magnetized neutron stars}
The oscillation equations for axisymmetric toroidal modes in the solid crust are given by
\begin{eqnarray}
r\frac{{\rm{d}}}{{\rm{d}} r}\pmb{t}=-\frac{1}{2}\frac{{\rm{d}}\ln
  a_1}{{\rm{d}}\ln r}\pmbmt{T}_A\pmb{t}+\pi
e^{\Lambda}\pmbmt{T}_B\pmb{W},
\end{eqnarray}
\begin{eqnarray}
r\frac{{\rm{d}}}{{\rm{d}} r}\pmb{W}=-\frac{e^{\Lambda}}{\beta}\pmbmt{K}_A\mbox{\boldmath
  $t$}-e^{\Lambda}\pmbmt{K}_B\pmb{W},
\end{eqnarray}
and those in the fluid regions by
\begin{eqnarray}
r\frac{{\rm{d}}}{{\rm{d}} r}\pmb{t}=-\frac{1}{2}\frac{{\rm{d}}\ln
  a_1}{{\rm{d}}\ln r}\pmbmt{T}_C\pmb{t}+\pi
e^{\Lambda}\pmbmt{T}_D\pmb{h},
\end{eqnarray}
\begin{eqnarray}
r\frac{{\rm{d}}}{{\rm{d}} r}\pmb{h}=-e^{\Lambda}\pmbmt{K}_C\pmb{t}-e^{\Lambda}\pmbmt{K}_D\pmb{h},
\end{eqnarray}
where the vectors $\pmb{W}$ and $\pmb{h}$ are defined, using the vector variables $\pmb{t}=\left(T_{l_j'}\right)$ and $\pmb{b}=\left(b_{l_j}\right)$, as
\begin{eqnarray}
\pmb{W}=\frac{\alpha}{\beta}e^{-\Lambda}\pmbmt{C}_0r\frac{{\rm{d}}}{{\rm{d}}
  r}\pmb{t}+\pmbmt{Q}_0\pmbmt{C}_1\pmb{h}, 
\end{eqnarray}
\begin{eqnarray}
\pmb{h}=\frac{r^3}{\sqrt{\pi}a_1}\pmb{b}.
\end{eqnarray}
Here, the matrices in the differential equations are defined by
\begin{eqnarray}
\pmbmt{T}_A=\pmbmt{M}^{-1}\pmbmt{S}_1,
\end{eqnarray}
\begin{eqnarray}
\pmbmt{T}_B=\pmbmt{M}^{-1}\pmbmt{C}_1\left(\pmbmt{Q}_0\pmbmt{C}_1\right)^{-1},
\end{eqnarray}
\begin{eqnarray}
\pmbmt{T}_C=\left(\pmbmt{Q}_1\pmbmt{C}_0\right)^{-1}\pmbmt{S}_1,
\end{eqnarray}
\begin{eqnarray}
\pmbmt{T}_D=\left(\pmbmt{Q}_1\pmbmt{C}_0\right)^{-1}\pmbmt{C}_1,
\end{eqnarray}
\begin{eqnarray}
\pmbmt{K}_A=e^{-2\Phi}Vc_1\bar{\omega}^2\bigg[\left(1+\frac{p}{\rho}+\frac{p}{\rho}\frac{\beta}{\pi}\eta\right)\pmbmt{C}_0
+\frac{p}{\rho}\frac{\beta}{\pi}\left(1-\eta\right)\pmbmt{Q}_0\pmbmt{Q}_1\pmbmt{C}_0\bigg]
\nonumber \\
+\alpha\left(2\pmbmt{C}_0-\pmbmt{C}_0\pmbmt{\Lambda}_1\right)
+\alpha\eta\pmbmt{S}_A\pmbmt{T}_A,
\end{eqnarray}
\begin{eqnarray}
\pmbmt{K}_B=e^{-\Lambda}\left(r\frac{{\rm{d}}\Phi}{{\rm{d}}r}+2\frac{{\rm{d}}\ln
    a_1}{{\rm{d}}\ln
    r}-1\right)\pmbmt{I}
+\frac{1}{2}e^{-\Lambda}\frac{{\rm{d}}\ln
  a_1}{{\rm{d}}\ln r}\left(\pmbmt{S}_B-\frac{\pi\alpha}{\beta}\pmbmt{S}_A\pmbmt{T}_B\right),
\end{eqnarray}
\begin{eqnarray}
\pmbmt{K}_C=\frac{e^{-2\Phi}Vc_1\bar{\omega}^2}{\beta}\bigg[\left(1+\frac{p}{\rho}+\frac{p}{\rho}\frac{\beta}{\pi}\eta\right)\left(\pmbmt{Q}_0\pmbmt{C}_1\right)^{-1}\pmbmt{C}_0
\nonumber \\
+\frac{p}{\rho}\frac{\beta}{\pi}\left(1-\eta\right)\left(\pmbmt{Q}_0\pmbmt{C}_1\right)^{-1}\pmbmt{Q}_0\pmbmt{Q}_1\pmbmt{C}_0\bigg],
\end{eqnarray}
\begin{eqnarray}
\pmbmt{K}_D=e^{-\Lambda}\left(r\frac{{\rm{d}}\Phi}{r}+2\frac{{\rm{d}}\ln
    a_1}{{\rm{d}}\ln r}-1\right)\pmbmt{I}
+\frac{1}{2}e^{-\Lambda}\frac{{\rm{d}}\ln a_1}{{\rm{d}}\ln r}\left(\pmbmt{Q}_0\pmbmt{C}_1\right)^{-1}\pmbmt{S}_0,
\end{eqnarray}
\begin{eqnarray}
\pmbmt{S}_A=\pmbmt{S}_0\left(\pmbmt{Q}_0\pmbmt{C}_1\right)^{-1}\pmbmt{C}_0,
\end{eqnarray}
\begin{eqnarray}
\pmbmt{S}_B=\pmbmt{S}_0\left(\pmbmt{Q}_0\pmbmt{C}_1\right)^{-1},
\end{eqnarray}
\begin{eqnarray}
\pmbmt{S}_0=\pmbmt{\Lambda}_0-\pmbmt{Q}_0\pmbmt{Q}_1\pmbmt{\Lambda}_0-2\pmbmt{Q}_0\pmbmt{C}_1,
\end{eqnarray}
\begin{eqnarray}
\pmbmt{S}_1=\pmbmt{\Lambda}_1-\pmbmt{Q}_1\pmbmt{Q}_0\pmbmt{\Lambda}_1+2\pmbmt{Q}_1\pmbmt{C}_0,
\end{eqnarray}
\begin{eqnarray}
\pmbmt{M}=\pmbmt{Q}_1\pmbmt{C}_0+\frac{\pi\alpha}{\beta}\pmbmt{C}_1\left(\pmbmt{Q}_0\pmbmt{C}_1\right)^{-1}\pmbmt{C}_0,
\end{eqnarray}
$\pmbmt{I}$ is the unit matrix, and
\begin{eqnarray}
\bar{\omega}=\frac{\omega}{\sqrt{M/R^3}} , \quad V=\frac{\rho gr}{p}, \quad
 g=\frac{M_r}{r^2}, \quad
c_1=\frac{\left(r/R\right)^3}{M_r/M}, \quad \alpha=\frac{\mu}{p}, \quad \beta=\frac{a_1^2/r^4}{p},
\end{eqnarray}
where $M$ and $R$ are the gravitational mass and the radius of the star, respectively. 
Non-zero elements of the matrices $\pmbmt{C}_0$, $\pmbmt{C}_1$, $\pmbmt{Q}_0$, $\pmbmt{Q}_1$, $\pmbmt{\Lambda}_0$, $\pmbmt{\Lambda}_1$ are
\begin{eqnarray}
\left(\pmbmt{C}_0\right)_{jj}=-\left(l_j+2\right)J_{l_j+1}^m, \quad 
\left(\pmbmt{C}_0\right)_{j+1,j}=\left(l_j+1\right)J_{l_j+2}^m, \nonumber
\end{eqnarray}
\begin{eqnarray}
\left(\pmbmt{C}_1\right)_{jj}=l_jJ_{l_j+1}^m, \quad \quad \quad \quad
\ \
\left(\pmbmt{C}_1\right)_{j,j+1}=-\left(l_j+3\right)J_{l_j+2}^m, \nonumber
\end{eqnarray}
\begin{eqnarray}
\left(\pmbmt{Q}_0\right)_{jj}=J_{l_j+1}^m, \quad \quad \quad \quad \quad
\left(\pmbmt{Q}_0\right)_{j+1,j}=J_{l_j+2}^m, \nonumber
\end{eqnarray}
\begin{eqnarray}
\left(\pmbmt{Q}_1\right)_{jj}=J_{l_j+1}^m, \quad \quad \quad \quad \quad
\left(\pmbmt{Q}_1\right)_{j,j+1}=J_{l_j+2}^m, \nonumber
\end{eqnarray}
\begin{eqnarray}
\left(\pmbmt{\Lambda}_0\right)_{jj}=l_j\left(l_j+1\right), \quad \quad
\quad \ 
\left(\pmbmt{\Lambda}_1\right)_{jj}=\left(l_j+1\right)\left(l_j+2\right) \nonumber
\end{eqnarray}
\begin{equation}
\end{equation}
for even modes, and
\begin{eqnarray} 
\left(\pmbmt{C}_0\right)_{jj}=l_jJ_{l_j+1}^m, \quad \quad \quad \quad
\ \
\left(\pmbmt{C}_0\right)_{j,j+1}=-\left(l_j+3\right)J_{l_j+2}^m,
\nonumber
\end{eqnarray}
\begin{eqnarray}
\left(\pmbmt{C}_1\right)_{jj}=-\left(l_j+2\right)J_{l_j+1}^m, \quad
\left(\pmbmt{C}_1\right)_{j+1,j}=\left(l_j+1\right)J_{l_j+2}^m,
\nonumber
\end{eqnarray}
\begin{eqnarray}
\left(\pmbmt{Q}_0\right)_{jj}=J_{l_j+1}^m, \quad \quad \quad \quad \quad
\left(\pmbmt{Q}_0\right)_{j,j+1}=J_{l_j+2}^m, \nonumber
\end{eqnarray}
\begin{eqnarray}
\left(\pmbmt{Q}_1\right)_{jj}=J_{l_j+1}^m, \quad \quad \quad \quad \quad
\left(\pmbmt{Q}_1\right)_{j+1,j}=J_{l_j+2}^m, \nonumber
\end{eqnarray}
\begin{eqnarray}
\left(\pmbmt{\Lambda}_0\right)_{jj}=\left(l_j+1\right)\left(l_j+2\right)
\quad \ 
\left(\pmbmt{\Lambda}_1\right)_{jj}=l_j\left(l_j+1\right), \nonumber
\end{eqnarray}
\begin{equation}
\end{equation}
for odd modes, where
\begin{eqnarray}
J_l^m=\sqrt{\frac{l^2-m^2}{4l^2-1}}.
\end{eqnarray}

\end{document}